\documentclass[onecolumn]{aa}
\usepackage{savesym}

\savesymbol{tablenum}
\usepackage[output-decimal-marker={.}, exponent-product=\cdot, per-mode=fraction]{siunitx}
\restoresymbol{SIX}{tablenum}

\usepackage{graphicx}
\graphicspath{{figures/}}

\usepackage{amsmath, amssymb, bm}
\usepackage{tabularx}
\usepackage{multirow}
\usepackage{wrapfig}
\DeclareSIUnit[]\rsun
{\text{R\ensuremath{_{\textup{sun}}}}}

\usepackage{soul,xcolor}
\setstcolor{red}

\begin{document}

\title{How the area of solar coronal holes affects the properties of high-speed solar wind streams near Earth  - I. An analytical model}

\author{Stefan J. Hofmeister \inst{\ref{Graz},\ref{NY},\ref{AIP}}
\and Eleanna Asvestari \inst{\ref{Helsinki}} 
\and Jingnan Guo \inst{\ref{Kiel}, \ref{Hefei}}
\and Verena Heidrich-Meisner \inst{\ref{Kiel}}
\and Stephan G. Heinemann \inst{\ref{MPS}}
\and Jasmina Magdalenic \inst{\ref{Leuven}, \ref{ROB}}
\and Stefaan Poedts \inst{\ref{Leuven}, \ref{Lublin}}
\and Evangelia Samara \inst{\ref{Leuven}, \ref{ROB}}
\and Manuela Temmer \inst{\ref{Graz}}
\and Susanne Vennerstrom \inst{\ref{DTU}}
\and Astrid Veronig \inst{\ref{Graz}, \ref{Kanzelhohe}}
\and Bojan Vr\v{s}nak \inst{\ref{Hvar}}
\and Robert Wimmer-Schweingruber \inst{\ref{Kiel}}
}

\institute{Institute of Physics, University of Graz, Austria \label{Graz} 
\and Columbia Astrophysics Laboratory, Columbia University, New York, USA \label{NY} 
\and Leibniz Institut for Astrophysics Potsdam, Germany \label{AIP}
\and Department of Physics, University of Helsinki, Finland \label{Helsinki}
\and Institute of Experimental and Applied Physics, University of Kiel, Germany \label{Kiel}
\and CAS Key Laboratory of Geospace Environment, University of Science and Technology of China, Hefei, China \label{Hefei}
\and Max Planck Institute for Solar System Research, G\"ottingen, Germany \label{MPS}
\and Centre for mathematical Plasma Astrophysics (CmPA), Department of Mathematics, KU Leuven, Belgium \label{Leuven}
\and Solar-Terrestrial Centre of Excellence - SIDC, Royal Observatory of Belgium, Brussels, Belgium \label{ROB}
\and Institute of Physics, University of Maria Curie-Sk{\l}odowska, Lublin, Poland \label{Lublin}
\and DTU Space, Copenhagen, Denmark \label{DTU}
\and Kanzelh\"ohe Observatory for Solar and Environmental Research, University of Graz, Austria \label{Kanzelhohe}
\and Hvar Observatory, Faculty of Geodesy, University of Zagreb, Croatia \label{Hvar}
}

\date{Received <date> /
Accepted <date>}

\abstract{
Since the 1970s, it is empirically known that the area of solar coronal holes affects the properties of high-speed solar wind streams (HSSs) at Earth. We derive a simple analytical model for the propagation of HSSs from the Sun to Earth, and thereby show how the area of coronal holes and the size of their boundary regions affect the HSS velocity, temperature, and density near Earth.
We presume that velocity, temperature, and density profiles form across the HSS cross-section close to the Sun, and that these spatial profiles translate into corresponding temporal profiles in a given radial direction due to the solar rotation. These temporal distributions drive the stream interface to the preceding slow solar wind plasma, and disperse with distance from the Sun. The HSS properties at \SI{1}{AU} are then given by all HSS plasma parcels launched from the Sun that did not impinge into the stream interface at Earth distance.

We show that the velocity plateau region of HSSs as seen at \SI{1}{AU}, if apparent, originates from the center region of the HSS close to the Sun, whereas the velocity tail at \SI{1}{AU} originates from the trailing boundary region. The peak velocity of HSSs at Earth depends on the longitudinal width of the HSS close to the Sun. The shorter the longitudinal width of a HSS close to the Sun, the more of its 'fastest' HSS plasma parcels from the HSS core and trailing boundary region have impinged into the stream interface to the preceding slow solar wind, and the smaller is the peak velocity of the HSS at Earth. As the longitudinal width is statistically correlated to the area of coronal holes, this explains also the well-known empirical relationship between coronal hole areas and HSS peak velocities.   
Further, the temperature and density of HSS plasma parcels at Earth depend on their radial expansion from the Sun to Earth. The radial expansion is determined by the velocity gradient across the HSS boundary region close to the Sun, and gives the velocity-temperature and density-temperature relationships at Earth their specific shape. When considering a large number of HSSs, the presumed correlation between the HSS velocities and temperatures close to the Sun degrades only slightly up to \SI{1}{AU}, but the correlation between the velocities and densities is strongly disrupted up to \SI{1}{AU} due to the radial expansion. Finally, we show how the number of particles of the piled-up slow solar wind in the stream interaction region depends on the velocities and densities of the HSS and preceding slow solar wind plasma.
} 

\titlerunning{How the area of CHs affects the properties of HSSs near Earth}
\authorrunning{Hofmeister et al.}
\maketitle

\section{Introduction}
Since the 1970s, it is well-known that high-speed solar wind streams (HSSs), i.e., the fast, Alfv\'enic component of the solar wind, originate from solar coronal holes \citep{nolte1976}. Solar coronal holes are regions of reduced density and temperature compared to the ambient solar corona, and feature an open-to-interplanetary-space magnetic field topology. Along these open magnetic field lines, solar plasma is accelerated radially away from the rotating Sun, reaching supersonic speeds forming the large-scale HSSs that transcend our solar system.

The acceleration process of HSSs, and in general of the solar wind, is one of the large unsolved problems in solar physics \citep{cranmer2002}. There are two concurring theories for the acceleration. On the one hand, there are wave/turbulence driven models, which are based on the observation of slow magneto-acoustic waves traveling along the magnetic field lines upwards to larger heights \citep{banerjee2009}, and their turbulent dissipation by interaction between outgoing waves and incoming waves reflected in the outer corona \citep[e.g., review by ][]{cranmer2009, hahn2012, hahn2013}. On the other hand, there are reconnection/loop-opening models, which are based on continuous magnetic reconnection processes, e.g., with ephemeral magnetic regions within coronal holes \citep{wang2020}. Although it is not clear yet which of the two processes provides most of the energy needed for the solar wind acceleration, we are sure that the main part of the acceleration takes place at heights \SI{<20}{\rsun}, i.e., very close to the Sun \citep{chashei2005, wang2000, yakovlev2019}.

The direction of propagation of HSSs throughout the heliosphere is set at similar distances close to the Sun. Since up to the Alfv\'enic point, which is probably located at about \SIrange{12}{17}{\rsun} \citep{pizzo1983, marsch1984}, the magnetic energy density dominates both the thermal and kinetic energy density of HSSs. Thus, HSSs mainly follow the global magnetic field configuration at these heights.
At the Alfv\'enic point, the magnetic and kinetic energy density are equal. The magnetic field configuration gets affected by the outflowing solar wind, and consequently, magnetic field lines get stretched into the radial direction away from the Sun \citep{pneuman1971}. 

Starting from the Alfv\'enic point, the kinetic energy density becomes dominant, and the solar wind plasma mainly propagates radially away from the rotating Sun.
 Since the magnetic field is frozen into the solar wind but  also anchored in the rotating Sun, the heliospheric magnetic field winds up and forms an Archimedean spiral \citep{parker1958}. Different coronal sources of the various types of solar wind thereby correspond to different spiral sectors in the Archimedean spiral. Consequently, also the interface between HSS plasma and preceding ambient slow solar wind plasma forms a spiral. At the back of this interface, the faster HSS plasma impinges into the stream interface and gets thermalized, whereas in the front of this interface, preceding slow solar wind plasma gets piled up, forming a dense, hot stream interaction region (SIR) around the stream interface. 
 
 At Earth distance, HSSs usually appear as a gradual increase of velocity across the SIR, an occasionally apparent extended velocity plateau, and a slow decline of velocities back to slow solar wind velocities. \citet{nolte1977} studied eight HSSs in 1973 at Earth distance and found that most of the HSS plasma during its declining phase seen at Earth originated from a boundary region at the trailing edge of the coronal hole, having a width of \SI{\approx 6}{\degree}. This approximate size of the boundary regions has further been confirmed by a fast-latitude scan of Ulysses, finding a boundary length of \SI{\approx 6}{\degree} for the equator-ward boundary of a HSS originating from a polar coronal hole \citep{mccomas1998}. The velocities, temperatures, and ion charge states of undisturbed HSS plasma parcels are generally well correlated to each other, while their densities show a weak anti-correlation to the velocities \citep{neugebauer1966, burlaga1970a, burlaga1970b, ogilvie1985, xu2015}.  
 
Besides the above-mentioned correlations, there are two well-known relationships between the peak velocity of HSSs near Earth and decisive properties of their source coronal holes: 1) the relationship to the flux tube expansion factor of coronal holes \citep{wang1990}, and 2) to the area of coronal holes \citep{nolte1976}. The relationship to the flux tube expansion factor is usually interpreted in terms of at which height energy is deposited to accelerate the solar wind plasma close to the Sun \citep{wang1990}, and therefore thought to be related to the main acceleration phase of HSSs. Consequently, it is often employed by solar coronal models to provide solar wind properties at a spherical shell at a distance of \SI{0.1}{AU} from the Sun. These solar wind properties are then used as inner boundary conditions to model the solar wind in the heliosphere, e.g., by time-dependent 3D magnetohydrodynamic (MHD) models like ENLIL \citep{odstrcil2003} and EUHFORIA \citep{pomoell2018}. 
 
The relationship of the HSS speed near Earth to the area of coronal holes has first been found by \citet{nolte1976}, and confirmed by \citet{robbins2006, vrsnak2007, abramenko2009, karachik2011, rotter2012, hofmeister2018, garton2018, heinemann2020}. The corresponding Pearson's correlation coefficients range from \numrange{0.4}{0.96}, depending on the periods the datasets cover. \citet{robbins2006} have indicated, that the slope of this linear relationship could depend on the latitude of the coronal hole. In contrary, \citet{hofmeister2018} have shown that it is rather the latitudinal separation angle between the center of mass of the coronal hole and the satellite taking the in-situ measurements, which affects the peak velocity we see in the ecliptic at \SI{1}{AU}. \citet{garton2018} have shown that the peak velocities of HSSs statistically saturate at around \SI{700}{km s^{-1}} for coronal holes with a longitudinal width $\SI{>70}{\degree}$, and is also well correlated with the longitudinal widths of the coronal holes. \citet{hofmeister2020} showed that the empirical relationship between the peak velocities of HSSs and the area of their source coronal holes is obtained in MHD simulations, which neglect the acceleration phase and purely model the propagation phase of HSSs in the heliosphere. This demonstrates that this empirical relationship describes the propagation phase of HSSs from the Sun to Earth. 

In this study, we derive a simplified theoretical model for the propagation of HSSs from the Sun to Earth, and show how the area of coronal holes affects the solar wind properties near Earth. This model is not self-consistent, but contains all relevant mechanisms to explain the empirical relationships between the HSS properties near Earth and the area of their source coronal holes. The study is divided in five main sections. In Section \ref{sect:assumptions}, we introduce the approximation that disturbances in HSSs only have local effects, such that disturbances generated at the HSS boundary regions only affect the boundary regions. This assumption also involves that plasma parcels launched from the HSS core region into a given radial direction travel at a constant speed through interplanetary space, before reaching the interface to the preceding slow solar wind. We refer to this approximation as the assumption of freely propagating HSSs. 
In Section \ref{sec:theory_freeprop}, we start at the Sun, presume that the cross section of HSSs close to the Sun is similar to the area of their source coronal holes, and consists of a core region and a boundary region. We derive how the properties across the cross section translate into temporal distributions into a given radial direction due to the solar rotation. Further, we derive how the plasma parcels evolve with distance from the Sun under the assumption of freely propagating HSSs. In Section \ref{sec:streaminterface}, we relax the assumption of freely propagating HSSs and add the stream interface to our calculations. We study the kinematics of the stream interface between the HSS and the preceding slow solar wind, using momentum conservation between the impinging HSS plasma at its back and piling-up slow solar wind plasma at its front. In Section \ref{sec:hss_in_heliosphere}, we finally postulate that HSSs observed consist of all HSS plasma parcels launched at the Sun that did not impinge into the stream interface yet. We derive the evolution of the main properties of HSSs with distance from the Sun, how their properties at \SI{1}{AU} depend on each other, and how they depend on the cross-section of the HSSs close to the Sun. We also show how the properties of the stream interface relate to the properties of the HSSs, and we deliberate the limitations of our model.  In Section \ref{sec:conclusions}, we discuss our findings and present our conclusions.

\section{Starting assumptions}  \label{sect:assumptions}

\begin{figure}[t!]
\centering
 \includegraphics[width=.9\textwidth]{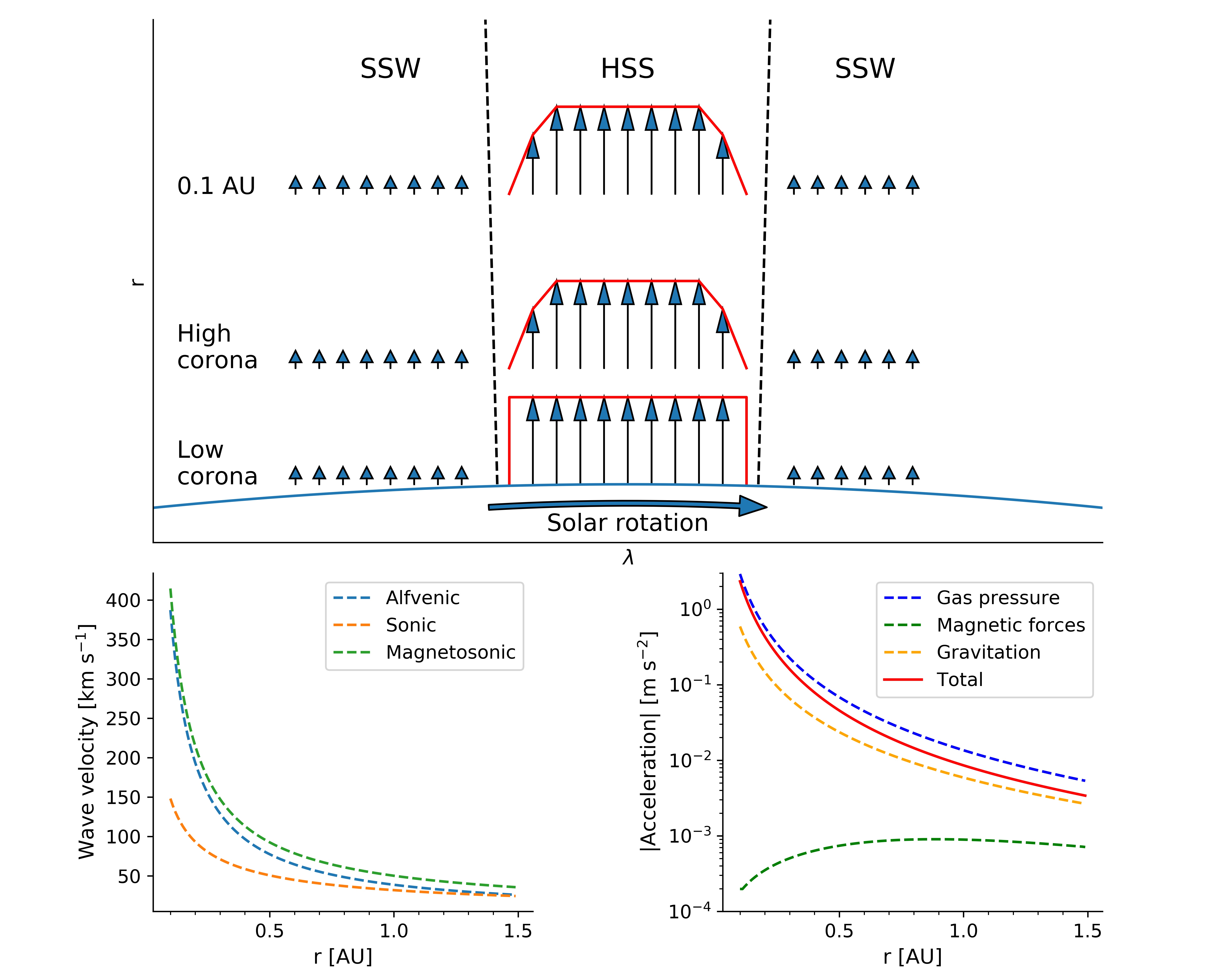}
\caption{Top: sketch of the formation of the velocity profile across the HSS cross section close to the Sun, which propagates in a channel within a slow solar wind environment. Velocities are delineated as arrows. Bottom left: acoustic, Alfv\'enic, and magneto-acoustic velocity in a HSS environment defined by Equations \ref{eq:theory_prophss0.1au1} and \ref{eq:theory_raddep}, versus its distance from the Sun. Bottom right: acceleration of a plasma parcel in a HSS due to gravitation and pressure gradients, as function of distance from the Sun.}
\label{fig:theory}
\end{figure}

In this section, we discuss the approximation of freely propagating HSS plasma parcels, which is an important concept needed in this study. This approximation implies that disturbances in HSSs only have local effects, and that therefore HSS plasma parcels launched into a given radial direction travel at a constant speed through interplanetary space as long as they do not reach the interface to the preceding slow solar wind.

In the following, we assume a HSS that has a circular cross-section at \SI{0.1}{AU}. Within the HSS, we assume an increased velocity $v_f$, temperature $T_f$, and decreased density $n_f$ as compared to the ambient slow solar wind stream properties $v_s$, $T_s$, and $n_s$. The magnetic flux density $B_f$ of the HSS is assumed to be the same as in the ambient slow solar wind $B_s$, as the expansion of the coronal hole is governed by the magnetic pressure which balances the magnetic flux densities. In MHD simulations, providing initial values of 
\begin{align}
    v_f = \SI{650}{km s^{-1}}, \quad n_f = \SI{150}{cm^{-3}}, 
    \quad T_f = \SI{1.6}{MK}, \quad B_f = \SI{216}{nT}
    \label{eq:theory_prophss0.1au1}
\end{align}
results in HSS velocities of \SI{730}{km s^{-1}}, densities of \SI{1.5}{cm^{-3}}, temperatures of \SI{75}{kK}, and magnetic field strength of about \SI{2}{nT} near Earth.
Initial values of
\begin{align}
    v_s = \SI{350}{km s^{-1}}, \quad n_s = \SI{500}{cm^{-3}}, 
    \quad T_s = \SI{400}{kK}, \quad B_s = \SI{216}{nT}
    \label{eq:theory_prophss0.1au2}
\end{align}
lead to slow solar wind velocities of \SI{350}{kms^{-1}}, densities of \SI{5}{cm^{-3}}, temperatures of \SI{20}{kK}, and magnetic field strength of about \SI{2}{nT} near Earth. Furthermore, the magnetic flux density chosen corresponds to a magnetic flux density of \SI{1}{G} in the low solar corona. We will use these values as our starting parameters for the HSSs and slow solar wind plasma at \SI{0.1}{AU}.

Further, throughout the study, we presume 1) that the HSS is a strongly compressible fluid consisting solely of protons, 2) that HSS plasma parcels propagate radially away from the rotating Sun with a constant velocity, 3) that the HSS undergoes adiabatic cooling, and 4) that the magnetic field  follows perfectly Parker's spiral. Within this section, we also approximate HSSs to be a steady-state system, which implies conservation of mass flux and magnetic flux within a solid angle as measured from the Sun; however, we will lift this restriction in the following sections, allowing the mass flux within a solid angle to change with distance from the Sun due to compression and dispersion effects.
For a stream propagating freely away from the Sun, i.e., experiencing no external forces such as interactions with other streams, these assumptions result in 
\begin{align}
    &v(r) = v_f, && B(r) = B_f \cdot \frac{r_0^2}{r^2}, && \hspace{3cm} \nonumber \\
    &n(r) = n_f \cdot \frac{r_0^2}{r^2}, && B_r(r) = B(r) \cdot \cos(\alpha), \nonumber \\
    &T(r) = T_f \cdot \left(\frac{r_0^2}{r^2}\right)^{\gamma -1}, &&  B_\lambda(r) =  B(r) \cdot \sin(\alpha), \nonumber \\
    &\alpha(r) = \arctan\left(\frac{r \cdot \omega}{v} \right). \label{eq:theory_raddep}
\end{align} 
Here, $r_0$ denotes the associated distance from the Sun at which $n_f$ and $T_f$ are defined, $\omega = 14^\circ/\si{d}$ is the average rotation rate of the Sun, $\gamma = 5/3$ is the adiabatic constant, $\alpha$ is the angle of the Parker spiral to the radial direction, $B_r$ is the component of the magnetic field strength in the radial direction, and $B_\lambda$ is the component perpendicular to the radial direction in the solar equatorial plane. In the following, we omit the explicit radial dependency of these variables whenever possible to improve readability. 

Next, we divide the propagation of HSSs from the Sun to Earth into different regimes. First, close to the Sun, we set a fast, low-density HSS in a slow, high-density solar wind environment. At the lateral boundaries of the flow, the HSS will be decelerated due to viscous forces and turbulence close to the Sun, resulting in a smooth transition of the velocity profile from the slow solar wind to HSS plasma, and consequently in a velocity profile over the cross-section of the HSS at \SI{0.1}{AU} (Fig. \ref{fig:theory} top).

The maximum speed with which any disturbances generated at the lateral boundary can travel from the boundary to the center of the stream is given by the maximum velocity of fast magneto-sonic waves $v_{ms}$, given by 
\begin{equation}  
    v_{ms}(r) = \sqrt{v_{so}^2 + v_a^2}, \hspace{1cm}
    v_{so}(r) = \sqrt{\frac{\gamma \ k_B \ T}{m_p}}, \hspace{1cm}
    v_{a}(r) = \frac{B}{\sqrt{\mu_0 \ n \ m_p}}, 
\end{equation}
where $v_{so}$ is the sonic speed, $v_{a}$ the Alfv\'en speed,  $m_p$ the proton mass, $\mu_0$ the vacuum permeability, and $k_B$ the Boltzmann constant. Using the HSS and slow solar wind properties defined above, we find that close to the Sun at \SI{0.1}{AU} the magneto-sonic speed is \SI{415}{km s^{-1}}, and strongly reduces to \SI{93}{km s^{-1}} at \SI{0.5}{AU} and \SI{50}{km s^{-1}} at \SI{1}{AU} (Fig. \ref{fig:theory} bottom left). Thus, disturbances generated at the boundaries close to the Sun can travel to and affect large parts of the HSS close to the Sun. With increasing distance from the Sun, the wave velocity decreases so that that these disturbances are mostly advected with the HSS, affecting only the local plasma conditions very close to the HSS boundary.  
Therefore, we may assume that the velocity profile of HSSs is formed mostly close to the Sun, and that with increasing distance from the Sun, the HSS plasma propagates freely in interplanetary space not seeing boundary effects. 

The radial acceleration of a freely propagating plasma parcel within the HSS environment can be approximated by the equation of motion, combined with the radial scaling laws for the plasma density, temperature, and magnetic field strength (Equ. \ref{eq:theory_raddep}). Assuming that the plasma parcel has the same density as the HSS, and with $m_p$ being the proton mass, $G$ being the gravitational constant, and $P$ being the thermodynamic pressure, the acceleration of such a plasma parcel is given by
\begin{align}
    \frac{dv}{dt} &= \left(-\nabla P - \nabla \frac{B^2}{2\ \mu_0} + \frac{(B \cdot \nabla) B}{\mu_0} - \frac{G \ m_p\ n \ M_\text{Sun}}{r^2}\right) / (m_p\ n)  \nonumber \\
    &= 2\ \gamma\ k_B \ m_p^{-1} \ T_f \frac{r_0^{2 \gamma -2}}{r^{2\gamma -1}} + \frac{B_f^2}{\mu_0\ m_p\ n_f} \frac{r_0^2}{r^3} \left(\sin^2(\alpha) - r\ \sin\alpha \ \cos\alpha\ \frac{d\alpha}{dr}  \right) - \frac{G \ M_\text{Sun}}{r^2}. \label{eq:theory_acceleration}
\end{align}

In the bottom right panel of Figure \ref{fig:theory}, we plotted the acceleration of such plasma parcels versus the distance from the Sun. The acceleration due to the radial gas pressure gradient, which decreases with $r^{2 (\gamma -1)} \approx r^{2.3}$, clearly dominates the total acceleration, with a minor contribution of the gravitation, and a negligible contribution due to the magnetic forces. 
The total acceleration due to the pressure gradients exceed the deceleration due to the gravitation at least until \SI{1.5}{AU}, and consequently the plasma is accelerated at least to this distance from the Sun. However, since the acceleration decreases with $\approx r^{2.3}$, the acceleration falls  below \SI{0.5}{m s^{-2}} already at a distance of \SI{0.2}{AU}. The associated total velocity increase of a plasma parcel having a velocity of \SI{650}{km s^{-1}} from \SIrange{0.2}{1}{AU} is only about \SI{15}{km s^{-1}}. Therefore, we neglect the radial acceleration of plasma parcels with distance as long as the plasma parcel did not reach the preceding SIR.

\section{Freely propagating HSS} \label{sec:theory_freeprop}

\begin{figure}[t!]
\centering
 \includegraphics[width=.9\textwidth]{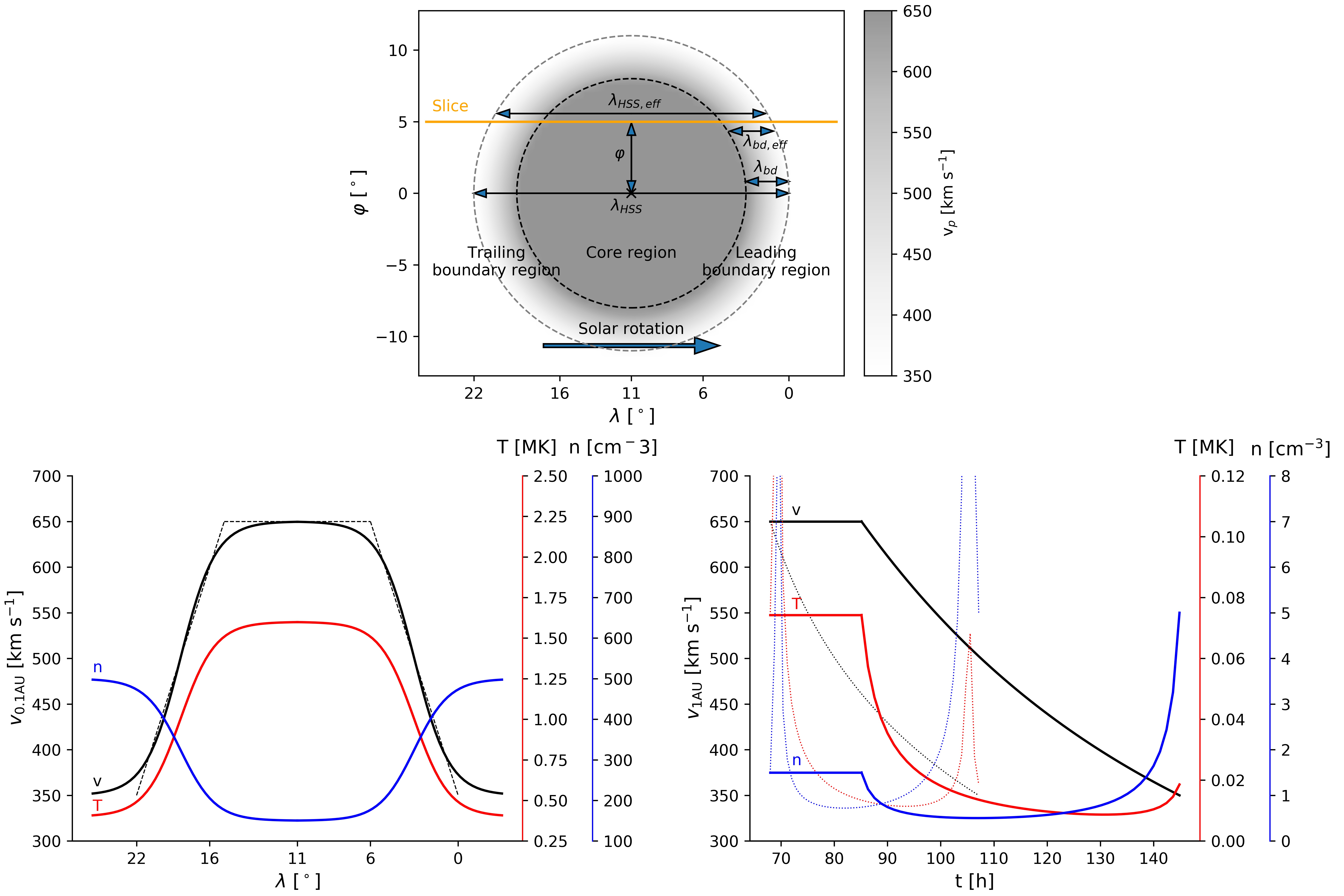}
\caption{Top: sketch of the velocity distribution across the cross-section of the HSS at \SI{0.1}{AU}, having a longitudinal widths of \SI{22}{\degree} and a widths of the boundary region of \SI{6}{\degree}. Bottom left: velocity (black), temperature (red), and density distribution (blue) along the cross-section of the HSS at \SI{0.1}{AU}. The solid lines give the sigmoidal approximations, and the black dashed line the linear approximation for the velocity profile. Bottom right: dispersed temporal velocity distribution as seen at \SI{1}{AU}, and the associated temperature and density distributions. Note, that the dotted lines represent the mathematical solution for plasma originating from the leading boundary region at \SI{0.1}{AU} of the HSS. These plasma parcels will have impinged into the stream interface at \SI{1}{AU} and thus cannot be observed. }
\label{fig:theory21}
\end{figure}

In this section, we derive a simplified model for the propagation of freely propagating HSSs. We assume a circular coronal hole in the low solar corona, and that the open magnetic field of the coronal hole expands with a constant, arbitrary expansion factor to \SI{0.1}{AU}, defining the cross-section of the associated HSS. This cross-section at \SI{0.1}{AU} consists of a nearly homogeneous core region as characterized by its temperature, density, and radial velocity distribution, and an adjunct boundary region across which the HSS properties adjust to the properties of the ambient slow solar wind. The rotation of the Sun translates these spatial distributions of the properties along the latitudinal slice into temporal distributions at a given radial direction. From there on, we presume that the HSS propagates freely in interplanetary space. Under these assumptions, we derive the velocity distribution of HSSs at a given distance from the Sun by the dispersion of its initial temporal velocity distribution close to the Sun. Further, we derive the evolution of the density and temperature of the associated plasma parcels in interplanetary space by calculating their expansion and presuming adiabatic cooling.

We define the properties of the HSS at \SI{0.1}{AU} in the following way, as sketched in Figure \ref{fig:theory21} top.
 We denote the circular diameter of the HSS cross-section, which is also the longitudinal width of the HSS, as $\lambda_{HSS}$, and the width of the boundary region as $\lambda_{bd}$. Thereby, $\lambda_{bd}$ is the width over which the HSS properties change by \SI{85}{\percent} from HSS to slow solar wind properties as measured from the center of the boundary region. Throughout the study, we keep the width of the boundary region fixed at \SI{6}{\degree}. 
We presume that the properties across the HSS solely depend on the distance to the center of the boundary region $d$, given by the sigmoid function: 
\begin{equation}
    \begin{pmatrix} v_{\SI{0.1}{AU}}(d) \\  T_{\SI{0.1}{AU}}(d) \\ n_{\SI{0.1}{AU}}(d) \end{pmatrix} = 
    \begin{pmatrix} v_s \\ T_s \\ n_s \end{pmatrix} + \begin{pmatrix} v_f - v_s \\ T_f - T_s \\ n_f - n_s \end{pmatrix} \cdot \frac{1}{2}\ \left(1 + \tanh{ \left( 2.5\  \frac{d - \lambda_{bd}/2}{\lambda_{bd}}\right)} \right) .
\end{equation}
The tangens hyperbolicus function was chosen over other sigmoid functions for calculation reasons. 

The associated velocity, temperature, and density distributions along a longitudinal slice through the center of the HSS is shown in the bottom left panel of Figure \ref{fig:theory21}. Since, in our formulation, the velocity, density, and temperature at \SI{0.1}{AU} depend only on the distance to the HSS boundary, they are also naturally correlated. This correlation allows to determine two of these quantities by knowing the third, using the normalized quantities
\begin{equation}
    \frac{v_{\SI{0.1}{AU}}(d) - v_s}{v_f - v_s} = \frac{T_{\SI{0.1}{AU}}(d) - T_s}{T_f - T_s} = \frac{n_{\SI{0.1}{AU}}(d) - n_s}{n_f - n_s} \label{eq:vnT_corr} .
\end{equation}
This relationship will be used later in this study.

\subsection{Dispersion of the velocity distribution with distance }

\begin{table}[t!]
\centering
\begin{tabular}{l c l l}
    & & $\text{for } 0.5\ \lambda_{HSS} - \varphi \ge \lambda_{bd} $ & $\text{for } 0.5\ \lambda_{HSS} - \varphi < \lambda_{bd}$  \\ \hline
    $\lambda_{HSS,eff}$ & $=$  & $2\ \sqrt{\left( 0.5\ \lambda_{HSS}\right)^2 - \varphi^2}$ & $2\ \sqrt{\left( 0.5\ \lambda_{HSS}\right)^2 - \varphi^2}$ \\
    $\lambda_{bd,eff}$ & $=$ & $0.5\ \lambda_{HSS,eff} - \sqrt{(0.5\ \lambda_{HSS} - \lambda_{bd} )^2 - \varphi^2} $ & $0.5\ \lambda_{HSS, eff}$  \\   
    $v_{sl,max} $ & $=$ & $v_f$ & $	v_s +  \frac{v_f - v_s}{\lambda_{bd}} \left(0.5\ \lambda_{HSS} - \varphi \right)$  \\
	$\left| \frac{dv}{d\lambda} \right| $ & $=$  & $\frac{v_{sl,max} - v_s}{\lambda_{bd,eff}}$ & $\frac{v_{sl,max} - v_s}{\lambda_{bd,eff}} $ \\[10pt]
\end{tabular}
\caption{Geometrical relations between the effective longitudinal width of a HSS along a given longitudinal slice$\lambda_{HSS,eff}$, the associated effective width of the HSS boundary region $\lambda_\text{bd,eff}$, the peak velocity in the slice $v_{sl,max}$, the associated velocity gradient across the boundary region $\frac{dv}{d\lambda}$, and the overall longitudinal width of the HSS $\lambda_{HSS}$, the HSS boundary region $\lambda_{bd}$, the displacement of the slice $\varphi$ to the center of the HSS, the maximum velocity of the HSS $v_f$, and the velocity of the slow solar wind $v_s$. The left solutions correspond to slices which cross the HSS core region, and the right solution corresponds to slices which lie purely in the northern/southern HSS boundary region.}\label{table:geometric_relations}
\end{table}

Next, we derive the temporal HSS velocity distribution at a given distance from the Sun for freely propagating HSSs. We start with the spatial velocity distribution along a longitudinal slice through the HSS close to the Sun (Fig. \ref{fig:theory21} top and bottom left). Using linear approximations for the velocity increase across the leading, i.e., western, and trailing, i.e., eastern boundary region, this is given by
\begin{equation}
    v_{\SI{0.1}{AU}}(\lambda) =
    \begin{cases}
        v_s + \lambda \cdot \frac{dv}{d\lambda} & \text{for } 0 \le \lambda \le \lambda_\text{bd,eff}, \\
        v_{sl,max} & \text{for } \lambda_\text{bd,eff} < \lambda < (\lambda_{HSS,eff} - \lambda_\text{bd,eff}), \\
        v_s + (\lambda_{HSS,eff} - \lambda) \cdot \frac{dv}{d\lambda} & \text{for } (\lambda_{HSS,eff} - \lambda_\text{bd,eff}) \le \lambda \le \lambda_{HSS,eff}.
    \end{cases} ,
\end{equation}
where $\lambda$ is a location in the slice measured from the leading edge along the slice, $\lambda_{HSS,eff}$ is the effective longitudinal width of the slice in the HSS,  $v_{sl,max}$ is the peak velocity in the slice,  $\lambda_\text{bd,eff}$ is the effective width of the associated boundary region along the slice, and $\frac{dv}{d\lambda}$ is the velocity gradient across the effective boundary region. It is important to note that the effective longitudinal width decreases, the width of the effective boundary region increases, and the velocity gradient across the effective boundary region decreases with increasing latitudinal displacement $\varphi$ of the slice to the center of the HSS. Further, also the peak velocity in the slice is decreased if either the entire slice is located in the northern/southern boundary region, or if the HSS is so small that the entire HSS consists solely of boundary region plasma. The mathematical relations of these parameters to the latitudinal displacement of the slice, the overall longitudinal width of the HSS, and the width of the boundary region follow from geometrical considerations and are given in Table \ref{table:geometric_relations}.

As the Sun rotates, this longitudinal velocity distribution $v_{\SI{0.1}{AU}}(\lambda)$ along a longitudinal slice in the HSS maps into a temporal velocity distribution $v_{\SI{0.1}{AU}}(t)$ in given, fixed heliospheric radial direction. Using $\lambda \rightarrow t = \frac{\lambda}{\omega}$, this results in
\begin{equation}
v_{\SI{0.1}{AU}}(t) = \left.
    \begin{cases}
    v_s + t\ \omega\ \frac{dv}{d\lambda} & \text{for } 0 \le t\ \omega \le \lambda_\text{bd,eff}, \\
    v_{sl,max} & \text{for } \lambda_\text{bd} < t\ \omega < (\lambda_{HSS,eff} - \lambda_\text{bd,eff}), \\
    v_s +  (\lambda_{HSS,eff} - t\ \omega) \cdot \frac{dv}{d\lambda} & \text{for } (\lambda_{HSS,eff} - \lambda_\text{bd,eff}) \le t\ \omega \le \lambda_{HSS,eff}.
    \end{cases} \right. \label{eq:theory_temporalprofile} 
\end{equation}
This temporal velocity distribution disperses with distance from the Sun. As we presume a freely propagating HSS, the temporal velocity distribution at an arbitrary distance $r$ from the Sun is related to the temporal velocity distribution at \SI{0.1}{AU} by
\begin{equation}
v_{\SI{0.1}{AU}}(t) = v_r\left(t + \frac{r}{v_{\SI{0.1}{AU}}} \right) .
\end{equation} 
By substituting $t + \frac{r}{v_{\SI{0.1}{AU}}}$ with $\breve{t}$, this leads to the following dispersed velocity distributions of the leading boundary region, plateau region, and trailing boundary regions of HSSs at a given distance from the Sun.
\begin{itemize}
\item Leading boundary region
\begin{equation}
    v_{r}(\breve{t}) = 
    \begin{cases}
        0.5\cdot \left(v_s +  \breve{t}\ \omega\ \frac{dv}{d\lambda}  + \sqrt{\left(v_s + \breve{t}\ \omega\ \frac{dv}{d\lambda} \right)^2 -4\ r\ \omega\ \frac{dv}{d\lambda}}   \right)    & \text{for }  r  < v_s\ v_{sl,max}\ \left(\omega\ \frac{dv}{d\lambda}\right)^{-1} \\ & \quad \quad \land \,  \frac{r}{v_s} \le \breve{t} \le \left( \frac{r}{v_{sl,max}} + \frac{\lambda_{bd,eff}}{\omega} \right),\\
    
        0.5\cdot \left(v_s +  \breve{t}\ \omega\ \frac{dv}{d\lambda}  - \sqrt{\left(v_s + \breve{t}\ \omega\ \frac{dv}{d\lambda} \right)^2 -4\ r\ \omega\ \frac{dv}{d\lambda}}   \right)    & \text{for } r \ge v_s\ v_{sl,max}\ \left(\omega\ \frac{dv}{d\lambda}\right)^{-1}   \\ & \quad \quad \land \, \left(\frac{r}{v_{sl,max}} + \frac{\lambda_{bd,eff}}{\omega}\right) \le \breve{t} \le \frac{r}{v_s} .  \label{eq:theory_vp_atr1}
     \end{cases}
\end{equation}
Note, that the term $v_s +  \breve{t}\ \omega\ \frac{dv}{d\lambda}$ is associated with the velocity of plasma parcels originating from the leading HSS boundary region at \SI{0.1}{AU}, i.e., $v_{\SI{0.1}{AU}}(t)$. The first line is valid for close distances from the Sun, as long as $ r < v_s\ v_{sl,max}\ \left(\omega\ \frac{dv}{d\lambda}\right)^{-1}$, i.e., as long as the HSS plasma stemming from the edge to the core region did not overtake the HSS plasma originating from the western edge of the boundary region. The second line is valid for the complementary case, taking place at distances $ r \ge  v_s\ v_{sl,max}\ \left(\omega\ \frac{dv}{d\lambda}\right)^{-1}$, i.e., when the plasma from the edge to the core region did overtake the plasma from western boundary edge. Note, that the second line, which is marked as a black dotted line in Figure \ref{fig:theory21} bottom left, is purely a mathematical solution. These plasma parcels will have impinged into the stream interface to the preceding slow solar wind plasma, breaking down the assumption of freely propagating plasma parcels. 

\item{Plateau region} 
\begin{equation}
    v_r(\breve{t}) = v_{sl,max}  \quad \quad \text{for } \left(\frac{r}{v_{sl,max}} + \frac{\lambda_{bd,eff}}{\omega}\right) < \breve{t} < \left(\frac{r}{v_{sl,max}} + \frac{\lambda_{HSS,eff} - \lambda_{bd,eff}}{\omega_\text{sun}}\right).  \label{eq:theory_vp_atr2}
\end{equation}
This equation expresses that the peak velocity of the HSS plateau at any distance to the Sun is given by the maximum velocity in the slice, as long as one assumes freely propagating HSSs, i.e., neglecting compression effects near the preceding slow solar wind in the SIRs, and any acceleration effects.

\item{Trailing boundary region} 
\begin{align}
 v_r(\breve{t}) &=
        0.5\cdot \left(v_s +  \left(\lambda_{HSS,eff} - \breve{t}\ \omega\right)\ \frac{dv}{d\lambda}  + \sqrt{\left(v_s + \left(\lambda_{HSS,eff} - \breve{t}\ \omega\right)\ \frac{dv}{d\lambda}  \right)^2 +4\ r\ \omega\ \frac{dv}{d\lambda}   } \right) \nonumber \\
            & \quad \quad \quad \text{for }  \left( \frac{r}{v_{sl,max}} + \frac{\lambda_{HSS,eff} - \lambda_{bd,eff}}{\omega} \right)  \le \breve{t} \le \left( \frac{r}{v_s} + \frac{\lambda_{HSS,eff}}{\omega} \right) . 
 \label{eq:theory_vp_atr3}
\end{align}
Note, that the term $v_s +  \left(\lambda_{HSS,eff} - \breve{t}\ \omega\right)\ \frac{dv}{d\lambda}$ is associated with the velocity of plasma parcels originating from the trailing HSS boundary region at \SI{0.1}{AU}, i.e., $v_{\SI{0.1}{AU}}(t)$.
\end{itemize}

\subsection{Density and temperature of freely propagating HSS plasma parcels}

\begin{figure}[t!]
\centering
 \includegraphics[width=.9\textwidth]{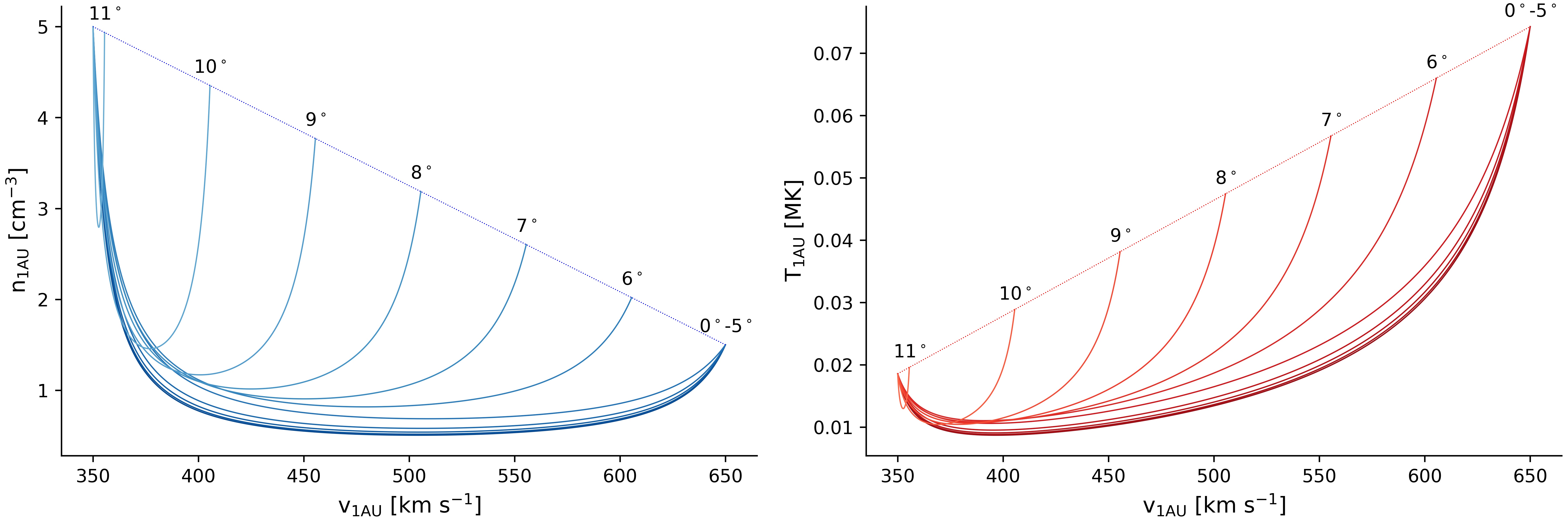}
\caption{Distribution of densities (left) and temperatures (right) of plasma parcels versus their velocities at \SI{1}{AU}, as derived by Equation \ref{eq:t_n_1AU} for a HSS having a diameter of \SI{22}{\degree}. Each line represents plasma parcels originating from a different longitudinal slice through the HSS, resulting in a different radial expansion. The numbers give the latitudinal displacement of the slice to the center of the HSS; slices \SI{>5}{\degree} lie entirely within the northern boundary region. For comparison, the dashed diagonal line shows the solution for the case that the radial expansion of plasma parcels would have been neglected.}
\label{fig:theory22}
\end{figure}

Next, we estimate the density and temperature of HSS plasma parcels at a distance $r$ from the Sun by deriving their expansion of volume with distance from the Sun, $\frac{V_{\SI{0.1}{AU}}(t)}{V_r(\breve{t})}$, and presuming adiabatic cooling,
\begin{align}
    &n_r(\breve{t}) = n_{\SI{0.1}{AU}}(t) \cdot \left(\frac{V_{\SI{0.1}{AU}}(t)}{V_r(\breve{t})}\right) \label{eq:Tn_r} \\
    &T_r(\breve{t}) = T_{\SI{0.1}{AU}}(t) \cdot \left(\frac{V_{\SI{0.1}{AU}}(t)}{V_r(\breve{t})}\right)^{\gamma -1} \label{eq:t_n_0.1au} 
\end{align}

The density and temperature of plasma parcels at \SI{0.1}{AU}, $n_{\SI{0.1}{AU}}$ and $T_{\SI{0.1}{AU}}$, can be derived using their correlation to the velocity at \SI{0.1}{AU} using Equation \ref{eq:vnT_corr}.
The expansion of the volume of plasma parcels propagating radially away from the Sun is given by their lateral and radial expansion. As we assume HSSs propagating radially away from the Sun, the lateral expansion follows an $r^2$ law. The radial expansion within plasma parcels can be determined by their intrinsic radial velocity gradient. This leads to the expansion of the volume 
\begin{equation}
    \frac{V_{\SI{0.1}{AU}}(t)}{V_r(\breve{t})} = \frac{r_0^2}{r_2} \cdot \left( 1 + \frac{dv_{\SI{0.1}{AU}}(t)}{dr} \cdot \frac{ r - \SI{0.1}{AU}}{v_{\SI{0.1}{AU}}(t)}\right)^{-1} .
\end{equation}

To evaluate this equation, a smooth function for the plasma velocity from the slow solar wind to HSS plasma is needed. Consequently, we approximate the velocity distribution across the leading and trailing HSS boundaries at \SI{0.1}{AU} again by sigmoidal functions, 
\begin{equation}
    v_{\SI{0.1}{AU}}(\lambda) =
    \begin{cases}
        v_s + \frac{v_{sl,max} - v_s}{2}\ \left(1 + \tanh{ \left( 2.5\  \frac{\lambda - \lambda_{bd,eff}/2}{\lambda_{bd,eff}}\right)} \right) & \text{for } \lambda < \lambda_{HSS,eff}/2 , \\
        v_s + \frac{v_{sl,max} - v_s}{2}\ \left(1 - \tanh{ \left( 2.5\  \frac{\lambda - \left( \lambda_{HSS,eff} -  \lambda_{bd,eff}/2 \right)}{\lambda_{bd,eff}}\right)} \right) & \text{for } \lambda > \lambda_{HSS,eff}/2 , 
    \end{cases}
\end{equation}
whereby the first line describes the leading HSS boundary region, and the second line the trailing HSS boundary region.
Taking the derivative leads to the increase of volume dependent on the distance from the Sun, given by:

\begin{equation}
    \frac{V_{\SI{0.1}{AU}}(t)}{V_r(\breve{t})} = \frac{r_0^2}{r^2} \cdot
    \begin{cases}
    \left( \left| 1 - \frac{5\ \omega\ \left( r - \SI{0.1}{AU} \right) \left( v_{\SI{0.1}{AU}}(t) - v_s \right) \left(v_{sl,max} - v_{\SI{0.1}{AU}}(t) \right)}{ \lambda_{bd,eff}\ v_{\SI{0.1}{AU}}(t)^2\ \left( v_{sl,max} - v_s \right) } \right|  \right) ^{-1}  & \text{for }  \frac{r}{v_s}  <  \breve{t} < \left( \frac{r}{v_{sl,max}} + \frac{\lambda_{HSS,eff} }{2\ \omega} \right) \\
     \left( 1 + \frac{5\ \omega\ \left( r - \SI{0.1}{AU} \right) \left(v_{\SI{0.1}{AU}}(t) - v_s \right) \left(v_{sl,max} - v_{\SI{0.1}{AU}}(t) \right)}{ \lambda_{bd,eff}\ v_{\SI{0.1}{AU}}(t)^2\ \left( v_{sl,max} - v_s \right) }  \right) ^{-1}  & \text{for } \left( \frac{r}{v_{sl,max}} + \frac{\lambda_{HSS,eff} }{2\ \omega} \right) < \breve{t} < \left( \frac{r}{v_s} + \frac{\lambda_{HSS,eff}}{\omega} \right) .
    \end{cases} \label{eq:t_n_1AU} .
\end{equation}
 
In the bottom right panel of Figure \ref{fig:theory21}, we have derived the temporal density and temperature behavior for the HSS of the top panel in Figure \ref{fig:theory21} at Earth distance. The dotted lines, which correspond to the leading edge of the HSS at \SI{0.1}{AU} and which are derived by the first line of Equation \ref{eq:t_n_1AU}, are purely mathematical solutions since these plasma parcels will have impinged into the stream interface. Following the plateau region of the HSS, the temperature decreases fast to slow solar wind values, whereas the density shows a small drop for an extended time, and then increases fast to slow solar wind values. Note, that this drop in density is an effect of the radial expansion of plasma parcels with distance from the Sun. It therefore depends on the chosen sigmoid function we used to model the velocity gradient across the boundary region and on the presumed boundary width. Using less steep sigmoid functions or wider boundary regions, respectively, will soften or even remove this density drop. In solar wind plasma measurements in the heliosphere, this density drop will only be observable if the HSS has a steep velocity gradient across its trailing boundary region close to the Sun, and if the HSS plateau region is very extended so that the reference density level before the drop is well-defined.

Further, in Figure \ref{fig:theory22}, we show the distribution of densities and temperatures of HSS plasma parcels at \SI{1}{AU} versus their velocities. These have been derived for a HSS having a longitudinal width of \SI{22}{\degree}, and slices through that HSS at latitudinal displacements in steps of \SI{1}{\degree} to the HSS center. For comparison, the dashed diagonal lines give the temperature and density distributions for the case of purely lateral expanding plasma parcels, i.e., that the radial expansion is neglected. At the endpoint of each line, i.e., $v_r = v_{sl,max}$, the plasma parcel originates from the center of the slice where the radial velocity gradient along the slice is $\approx \SI{0}{km s^{-1}}$. Therefore, the plasma parcels do not expand into the radial direction, and the solutions for considering and neglecting the radial expansion of plasma parcels with distance coincide. The same is valid for plasma parcels originating from pure slow solar wind plasma. For all other points along the lines, the plasma parcels originate from within the boundary region, where the radial velocity gradient along the slice is $\ne \SI{0}{km s^{-1}}$. These plasma parcels expand in the radial direction with distance from the Sun, additionally reducing their densities and temperatures compared to a purely lateral expansion.

\section{Kinematics of the stream interface} \label{sec:streaminterface}

\begin{figure}[t!]
\centering
 \includegraphics[width=.9\textwidth]{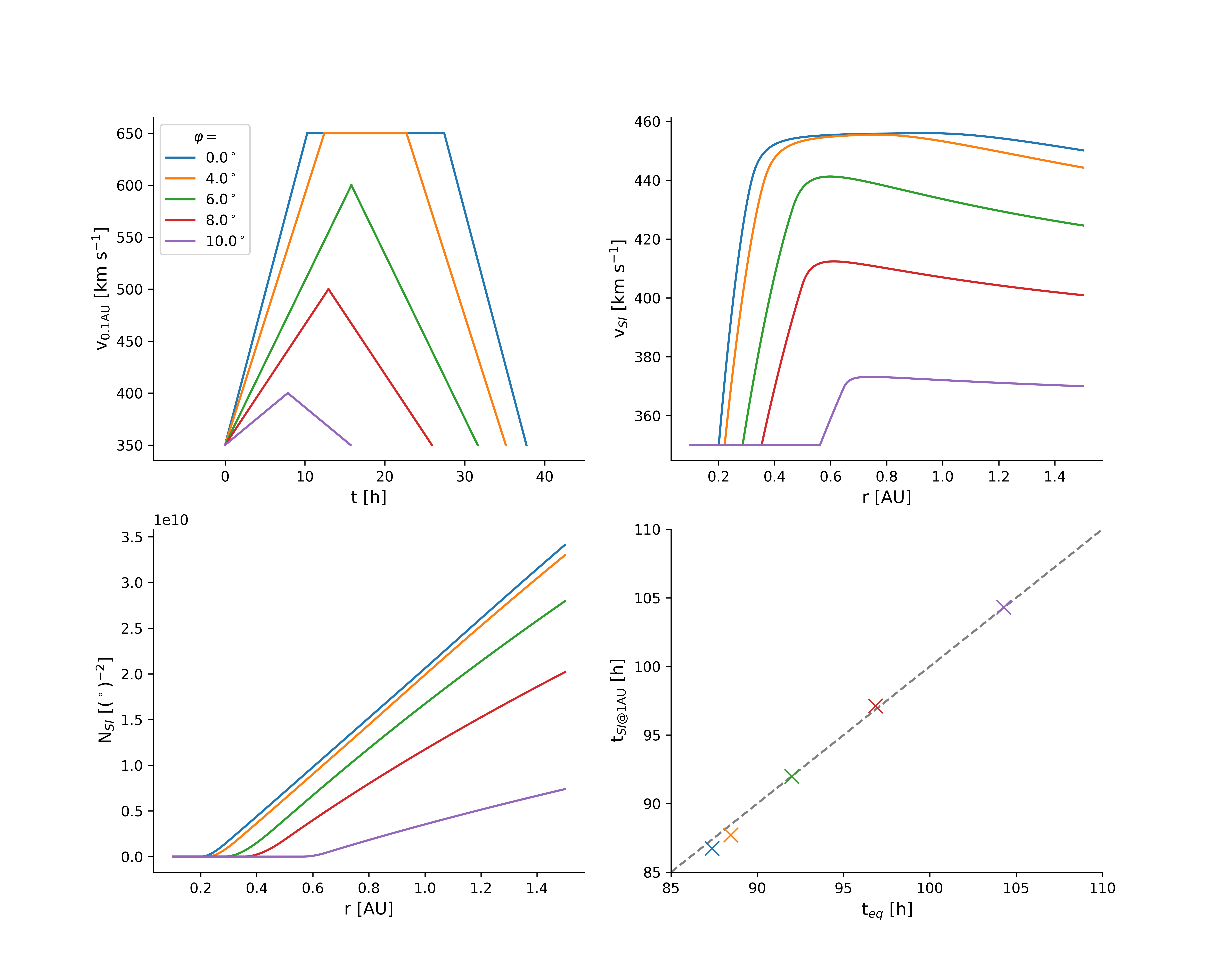}
\caption{Numerical solution of the kinematics of the stream interface. Top left: temporal velocity profile of the driving HSS plasma. The HSS is presumed to have a longitudinal width of \SI{22}{\degree}, and the slices are set at \SI{0}{\degree}, \SI{4}{\degree}, \SI{6}{\degree}, \SI{8}{\degree}, and \SI{10}{\degree} to the HSS center. Top right: resulting velocity of the stream interface, versus its distance from the Sun. Bottom left: total mass of the stream interface per solid angle, dependent on its distance from the Sun. Bottom right: numerical arrival time of the stream interface at \SI{1}{AU} versus the analytically approximated arrival time derived by Equation \ref{eq:si_arrival}.}
\label{fig:theory3}
\end{figure}

In this section, we derive the kinematics of the stream interface between the HSS and the preceding slow solar wind plasma, and estimate its arrival time at Earth. Due to the very high magnetic Reynolds numbers, the HSS and slow solar wind plasma are confined within the magnetic field and cannot mix; therefore, the HSS accumulates at the back of the stream interface, and the preceding slow solar wind is piled up at its front. By presuming the stream interface to be a discrete entity, we derive the kinematics of the stream interface by momentum conservation between the stream interface, the impinging HSS plasma at its back, and the piling-up slow solar wind plasma at its front.

At a given time $t$ and velocity of the stream interface $v_{SI}$, the number of accumulating HSS particles per time per area is given by $n_r(t)\ (v_r(t) - v_{SI})\ dt$, impinging with a velocity $(v_r(t) - v_{SI})$. Analogue, the number area density of the piling-up slow solar wind plasma is given by $n_s\ (v_{SI} - v_s)\ dt$ impinging with a velocity $(v_{SI} - v_s)$. With $N_{SI}$ being the number area density of the stream interface, the kinematics of the stream interface is therefore determined by
\begin{align}
    &\frac{dv_{SI}(t)}{dt} = \frac{1}{N_{SI}} \left( n_r\ (v_r - v_{SI})^2 - n_s\ (v_s - v_{SI})^2 \right) , \nonumber \\
     &\frac{dN_{SI}(t)}{dt} = n_s\ (v_{SI} - v_s) + n_r\ (v_r - v_{SI}). \label{eq:theory_si} 
\end{align}
When we presume constant properties of the HSS and slow solar wind plasma impinging into the stream interface per solid angle, the set of differential equations is solved by
\begin{align}
    & N_{SI}(t) = \sqrt{t^2\ n_r\ n_s\ \left(  v_r - v_s \right)^2 +2\ t\ N_0 \left( n_r\ (v_r - v_0) + n_s\ (v_0 - v_s) \right) + N_0^2},  \\
    & v_{SI}(t) = \frac{\frac{d}{dt}N_{SI}(t) + n_s\ v_s - n_r\ v_r}{n_s - n_r},
\end{align}
whereby $v_0$ and $N_0$ are the velocity and number area density of the stream interface at the time $t_0$.
When we further assume that the stream interface starts with a velocity of \SI{0}{km s^{-1}}, that the associated number area density related to a solid angle at the start is $0/(\si{\degree})^{2}$, and that the HSS velocity is equal to $v_{sl,max}$, the equations simplify. With $n_{sl,min}$ being the density at the center of the longitudinal slice within the HSS, which can be derived by Equation \ref{eq:vnT_corr}, the mass increase rate of the stream interface, its velocity, and its distance traveled at a time $t$ become
\begin{align}
    & \frac{d}{dt}N_{SI} = \sqrt{n_{sl,min}\ n_s} (v_{sl,max} - v_s), \\
    & v_{SI} = \frac{v_s\ \left(n_s - \sqrt{n_{sl,min}\ n_s}\right) +  v_{sl,max}\ \left(\sqrt{n_{sl,min}\ n_s} - n_{sl,min}\right)}{n_s - n_{sl,min}}, \\
    & r_{SI}(t) =  t\cdot v_{SI} \label{eq:theory_v_si}.
\end{align}
Note, that the velocity of the stream interface is independent on time. This means that the velocity immediately adjusts so that the accumulated HSS mass in its back balances the piling-up slow solar wind mass at its front, and it is constant as long as the impinging HSS and piling-up SSW properties stay constant. 

Next, we split the travel time of the stream interface from the Sun to Earth into two regions. First, we estimate the time and distance from the Sun at which the HSS plasma overtakes the preceding slow solar wind plasma, i.e., at which the SIR is formed. Up to this time, the stream interface roughly propagates with the velocity of the preceding slow solar wind as no momentum has yet been transferred from the HSS to the stream interface. Within the first-order approximation of linearly increasing velocities across the leading boundary region, all plasma parcels originating from the leading boundary region impinge into the stream interface at nearly the same time. Evaluating this condition for plasma parcels originating from the western and eastern edge of the leading boundary region, i.e., 
\begin{equation}
v_s\ t_1 = v_{sl,max}\ (t_1 - \lambda_{bd,eff} / \omega) ,
\end{equation}
one finds for the impingement time $t_1$, and for the corresponding distance $r_1$ at which the SIR is formed,
\begin{align}
t_1 &= v_{sl,max}\ \left(\omega\ \frac{dv}{d\lambda} \right)^{-1} \label{eq:ar_ldedge_si}, \\
r_1 &= v_s\ v_{sl,max}\ \left(\omega\ \frac{dv}{d\lambda} \right)^{-1}.
\end{align}

From there on, we neglect the following acceleration phase of the stream interface, presume that it is instantaneously accelerated to its equilibrium velocity given by Equation \ref{eq:theory_v_si}, and approximate it to further propagate with this constant equilibrium velocity. The arrival time of the stream interface $t_{SI,r}$ at a distance $r$ is then given by the time at which the SIR is formed plus the remaining travel time of the stream interface at its equilibrium speed up to the distance $r$, i.e., 
\begin{align}
t_{SI,r} = v_{sl,max}\ \left(\omega\ \frac{dv}{d\lambda} \right)^{-1} + \frac{ r - v_s\ v_{sl,max}\ \left(\omega\ \frac{dv}{d\lambda} \right)^{-1} } {v_{SI}} \quad \quad \text{for } r > r_1. \label{eq:si_arrival}
\end{align}

In the following, we apply these equations to the HSS plasma defined in Section \ref{sec:theory_freeprop}. We presume that the HSS has longitudinal widths of \SI{22}{\degree} close to the Sun, and evaluate the HSSs and associated stream interface properties at latitudinal displacements of \SI{0}{\degree}, \SI{4}{\degree}, \SI{6}{\degree}, \SI{8}{\degree}, and \SI{10}{\degree} to the HSS center. 
First, we derive the numerical solution for the kinematics of the stream interface, by using its defining set of differential equations combined with the full time-dependent solution of the impinging HSS plasma. For each time step, we derive iterative the HSS properties velocity and density at the location of the stream interface using Equations \ref{eq:theory_vp_atr1} to \ref{eq:theory_vp_atr3} and Equation \ref{eq:t_n_1AU}, assuming the preceding slow solar wind to have constant properties defined by Equations \ref{eq:theory_prophss0.1au1} and \ref{eq:theory_prophss0.1au2}, and update the velocity, number density, and location of the stream interface using Equations \ref{eq:theory_si}. In the top right panel of Figure \ref{fig:theory3}, we show the velocity of the stream interface versus its distance. Starting at the Sun, the velocity of the stream interface first coincides with the velocity of the preceding slow solar wind. Next, the velocity increases sharply to the steady-state solution given by Equations \ref{eq:theory_v_si}, which is related to the formation of the SIR by the impinging HSS plasma. Finally, the velocity of the stream interface slowly declines again, which is related to the reduced velocity of HSS plasma parcels originating from the trailing boundary region of the HSS. The much slower decrease of the velocity of the stream interface as compared to its sharp increase is explained by its larger, accumulated area density and momentum at larger distances, which is shown in the bottom left panel of Figure \ref{fig:theory3}. While the accumulated particles per area are close to zero near the Sun, they roughly increase linearly as soon as the SIR is formed.
In the bottom right panel of Figure \ref{fig:theory3}, we compare the estimated arrival times at \SI{1}{AU} derived by Equation \ref{eq:si_arrival} versus the exact arrival times derived by solving the differential Equation \ref{eq:theory_si} numerically. The  estimated arrival times match the exact arrival times within an error of \SI{1}{hour}, and are therefore good analytical proxies which are used in the next section.

\section{Characteristic properties of HSS and the stream interfaces in the inner heliosphere} \label{sec:hss_in_heliosphere}

In this section, we postulate that the HSS plasma observed by satellites consist of all HSS plasma parcels launched from the Sun, that did not impinge into the stream interface at the satellite's distance. Accordingly, we study the properties of HSSs by evaluating only freely propagating HSS plasma parcels, derived in Section \ref{sec:theory_freeprop}, that arrive later than the stream interface at the given distance, derived in Section \ref{sec:streaminterface}. 

We first derive the characteristic properties of HSSs, which we define as the properties of the fastest HSS plasma parcel, as a function of the distance from the Sun. In case that not the entire HSS core region has impinged into the stream interface, the velocity of the fastest parcel is equal to the velocity of the HSS core region. Else, the velocity is given by the fastest plasma parcel originating from the trailing boundary region that did not impinge into the stream interface yet. These cases can be approximated by freely propagating HSSs (Equ. \ref{eq:theory_vp_atr2} to \ref{eq:theory_vp_atr3}) evaluated at the arrival time of the stream interface at the given distance (Equ. \ref{eq:si_arrival}), which results in:
\begin{align}
v_p(r) &= \begin{cases}
v_{sl,max} &\text{for } t_{SI} \le \frac{\lambda_{HSS,eff} - \lambda_{bd,eff}}{\omega} + \frac{r}{v_{sl,max}}, \\
 0.5\ \left( v_s - v_{sl,max} + \frac{v_{sl,max}\ v_s}{v_{SI}} - \frac{r\ \omega}{v_{SI}}\ \frac{dv}{d\lambda} + \lambda_{HSS,eff}\ \frac{dv}{d\lambda} +  
 \vphantom{ \quad + \sqrt{\left( v_s - v_{sl,max} + \frac{v_{sl,max}\ v_s}{v_{SI}} - \frac{r\ \omega}{v_{SI}}\ \frac{dv}{d\lambda} + \lambda_{HSS,eff}\ \frac{dv}{d\lambda} \right)^2 + 4\ r\ \omega\ \frac{dv}{d\lambda}  } } \right. \\
  \left. \quad + \sqrt{\left( v_s - v_{sl,max} + \frac{v_{sl,max}\ v_s}{v_{SI}} - \frac{r\ \omega}{v_{SI}}\ \frac{dv}{d\lambda} + \lambda_{HSS,eff}\ \frac{dv}{d\lambda} \right)^2 + 4\ r\ \omega\ \frac{dv}{d\lambda}  } \right) &\text{for } t_{SI} >  \frac{\lambda_{HSS,eff} - \lambda_{bd,eff}}{\omega} + \frac{r}{v_{sl,max}} .
\end{cases} \label{eq:vat1au}
\end{align}

Therefore, the peak velocity is dependent on the properties of the slow solar wind and of the HSS, the velocity of the stream interface, the effective longitudinal width of the HSS, the velocity gradient across the HSS boundary close to the Sun, and the distance at which the equation is evaluated. Note, that the effective longitudinal widths of the HSS and the longitudinal velocity gradient across the HSS boundary are also implicitly dependent on the latitudinal position of the longitudinal slice within the HSS (Table \ref{table:geometric_relations}).

The temperature and density of the plasma parcels associated with the peak velocity can be determined by first deriving the plasma temperature and density of these parcels close to the Sun using the correlation between the velocity, temperature, and density, i.e., Equation \ref{eq:vnT_corr}, and then deriving their expansion and associated cooling by Equations \ref{eq:t_n_0.1au}, \ref{eq:Tn_r}, and \ref{eq:t_n_1AU}: 
\begin{align}
&n_{@vp}(r) =
n_{\SI{0.1}{AU}}(v_p) \cdot \left( \frac{\SI{0.1}{AU}}{r} \right)^2 \cdot \begin{cases} 1 &\text{for } t_{SI} \le \frac{\lambda_{HSS,eff} - \lambda_{bd,eff}}{\omega} + \frac{r}{v_{sl,max}}\\
\left( 1 + \frac{5\ \omega\ \left( r - \SI{0.1}{AU} \right) \left(v_p - v_s \right) \left(v_{sl,max} - v_p \right)}{ \lambda_{bd,eff}\ v_p^2\ \left( v_{sl,max} - v_s \right) }   \right)^{-1} \quad \quad \quad  &\text{for } t_{SI} >  \frac{\lambda_{HSS,eff} - \lambda_{bd,eff}}{\omega} + \frac{r}{v_{sl,max}}
\end{cases}, \label{eq:nTat1au1} \\
&T_{@vp}(r) = T_{\SI{0.1}{AU}}(v_p) \cdot \left( \frac{\SI{0.1}{AU}}{r} \right)^{2 (\gamma -1)} \cdot \begin{cases} 1 &\text{for } t_{SI} \le \frac{\lambda_{HSS,eff} - \lambda_{bd,eff}}{\omega} + \frac{r}{v_{sl,max}}, \\
\left( 1 + \frac{5\ \omega\ \left( r - \SI{0.1}{AU} \right) \left(v_p - v_s \right) \left(v_{sl,max} - v_p \right)}{ \lambda_{bd,eff}\ v_p^2\ \left( v_{sl,max} - v_s \right) }   \right)^{1 - \gamma}  &\text{for } t_{SI} >  \frac{\lambda_{HSS,eff} - \lambda_{bd,eff}}{\omega} + \frac{r}{v_{sl,max}} .
\end{cases}
\label{eq:nTat1au2}
\end{align}
In the next sections, we analyze the behavior of theses properties with distance from the Sun and their interrelations in more detail. 

\subsection{Evolution of the velocity, density and temperature with distance from the Sun}

\begin{figure}[t!]
\centering
 \includegraphics[width=.9\textwidth]{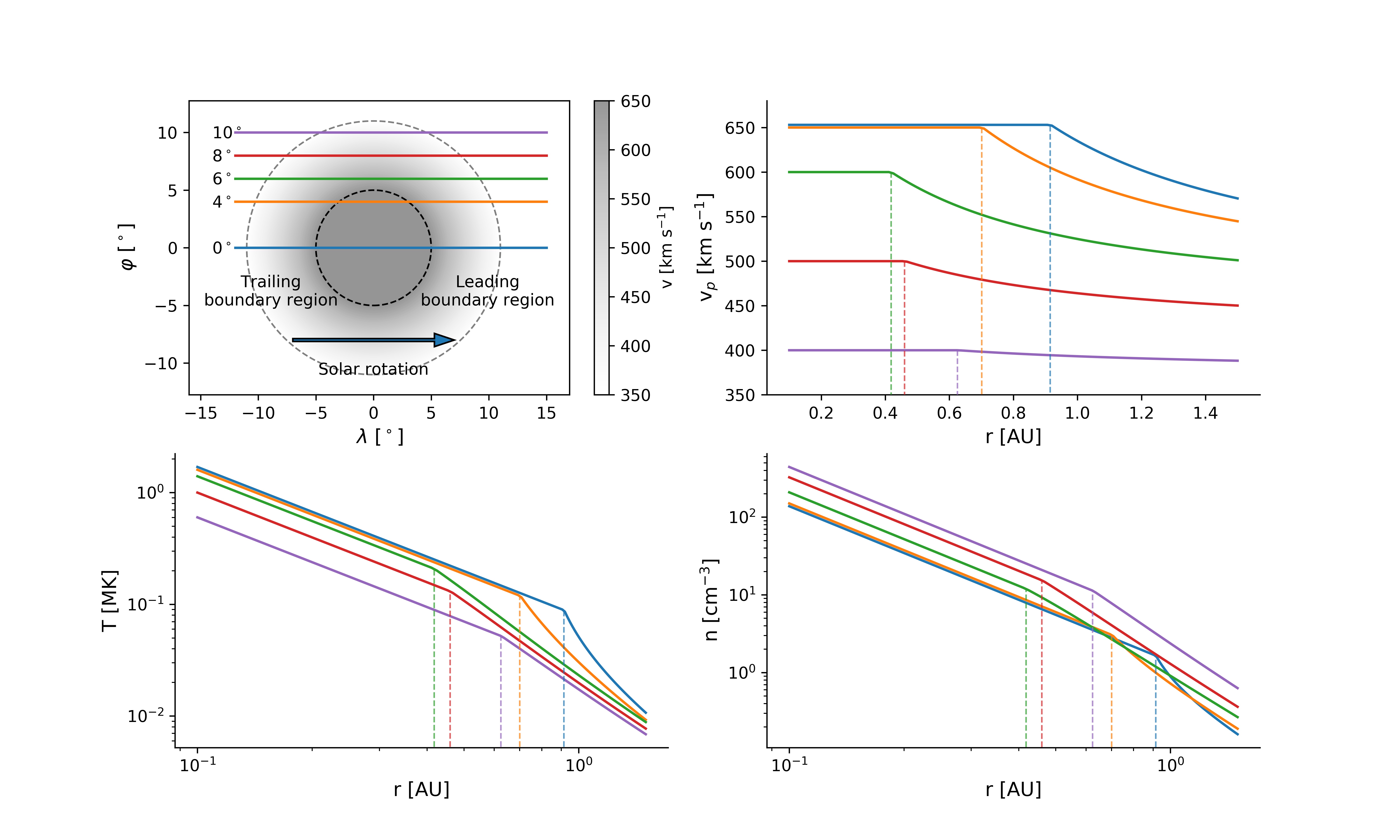}
\caption{Schematic cross-section of a HSS at \SI{0.1}{AU}, having a longitudinal widths of \SI{22}{\degree} (top left). Peak velocity of the HSS versus its distance from the Sun (top right). Temperature (bottom left) and density (bottom right) of the associated HSS plasma parcels versus the distance from the Sun. The HSS properties have been derived at latitudinal displacements of \SI{0}{\degree} (blue), \SI{4}{\degree} (orange), \SI{6}{\degree} (green), \SI{8}{\degree} (red), and \SI{10}{\degree} (purple) to the center of the HSS. The dashed vertical lines mark the distance at which the plasma parcels start to originate from the trailing boundary region.}
\label{fig:theory45}
\end{figure}

First, we present how the velocity, density, and temperature of HSSs, represented by the plasma parcels having the peak velocity, behave with distance from the Sun.
 We again presume a HSS having a longitudinal width of \SI{22}{\degree} close to the Sun, and evaluate Equations \ref{eq:vat1au}, \ref{eq:nTat1au1} and \ref{eq:nTat1au2} at latitudinal displacements of \SI{0}{\degree}, \SI{4}{\degree}, \SI{6}{\degree}, \SI{8}{\degree}, and \SI{10}{\degree} to the HSS center, as sketched in the top left panel of Figure \ref{fig:theory45}.
 
 In the top right panel of Figure \ref{fig:theory45}, we show the peak velocity of the HSS versus its distance from the Sun. In the center of the HSS (blue line), the peak velocity is constant with \SI{650}{km s^{-1}} up to a distance of \SI{0.92}{AU}, and decreases from thereon continuously. Therefore, up to \SI{0.92}{AU}, all these plasma parcels originate from the core region of the HSS. Starting from \SI{0.92}{AU}, the peak velocity is given by the fastest plasma parcel from the trailing boundary region, that did not impinge into the stream interface yet. As more `fastest' plasma parcels impinge into the stream interface, the peak velocity decreases with distance. The rate of decrease thereby mostly depends on the velocity gradient across the HSS boundary region close to the Sun, and the velocity with which the stream interface propagates away from the Sun (Eq. \ref{eq:vat1au}). 
At a latitudinal displacement of \SI{4}{\degree}, the longitudinal slice still crosses the HSS core region. Accordingly, the maximum velocity in the longitudinal slice, and thus the peak velocity of the HSS, is still \SI{650}{km s^{-1}}. However, the effective longitudinal width of the HSS is reduced, which results in an effective smaller HSS core region along the slice. A smaller core region is equivalent to less time until all plasma parcels of the core region have impinged into the stream interface, and thus an earlier onset of the decrease of the peak velocities with distance. 
At latitudinal displacements of \SI{6}{\degree}, \SI{8}{\degree} and \SI{10}{\degree}, the longitudinal slice through the HSS crosses only the boundary region of the HSS. This results in a reduced maximum velocity within the slice, and consequently in smaller peak velocities already close to the Sun. As the velocity gradient across the leading HSS boundary region is smaller at larger latitudinal displacements to the HSS center, the time by which HSS plasma parcels have caught up with the preceding slow solar wind, and thus the distance at which the SIR forms, increases (see Equ. \ref{eq:si_arrival}). Up to that distance, the HSS peak velocity is given by the maximum velocity within the longitudinal slice. Right afterwards, the peak velocities start to decrease as there is no HSS plateau region.

In the bottom panels of Figure \ref{fig:theory45}, we show the associated densities and temperatures of the plasma parcels representing the peak velocities versus the distance from the Sun, in a double-logarithmic plot. As long as all the plasma parcels originate either from the HSS core region or from the longitudinal center of the HSS, i.e., where the velocity gradient along the longitudinal slice at \SI{0.1}{AU} is zero, the plasma parcels only expand laterally. Consequently, the density decreases with $r^{-2}$ and the temperature with $r^{2 (\gamma -1)}$. As soon as the HSS peak velocity gets associated with plasma parcels originating from the trailing boundary region, for which the longitudinal velocity gradient across the boundary region is $\ne \SI{0}{km s^{-1}}$, the plasma parcels also expand radially. Due to the radial expansion, the densities of these plasma parcels decrease faster than $r^{-2}$, and the temperatures decrease faster than $r^{2 (\gamma -1)}$. Therefore, in general, it should not be assumed that the densities in HSSs decrease with $r^{-2}$, with the exception of plasma parcels originating from the HSS core region.

\subsection{Dependence of the properties of HSSs at \SI{1}{AU} on their geometry close to the Sun}

\begin{figure}[t!]
\centering
 \includegraphics[width=.9\textwidth]{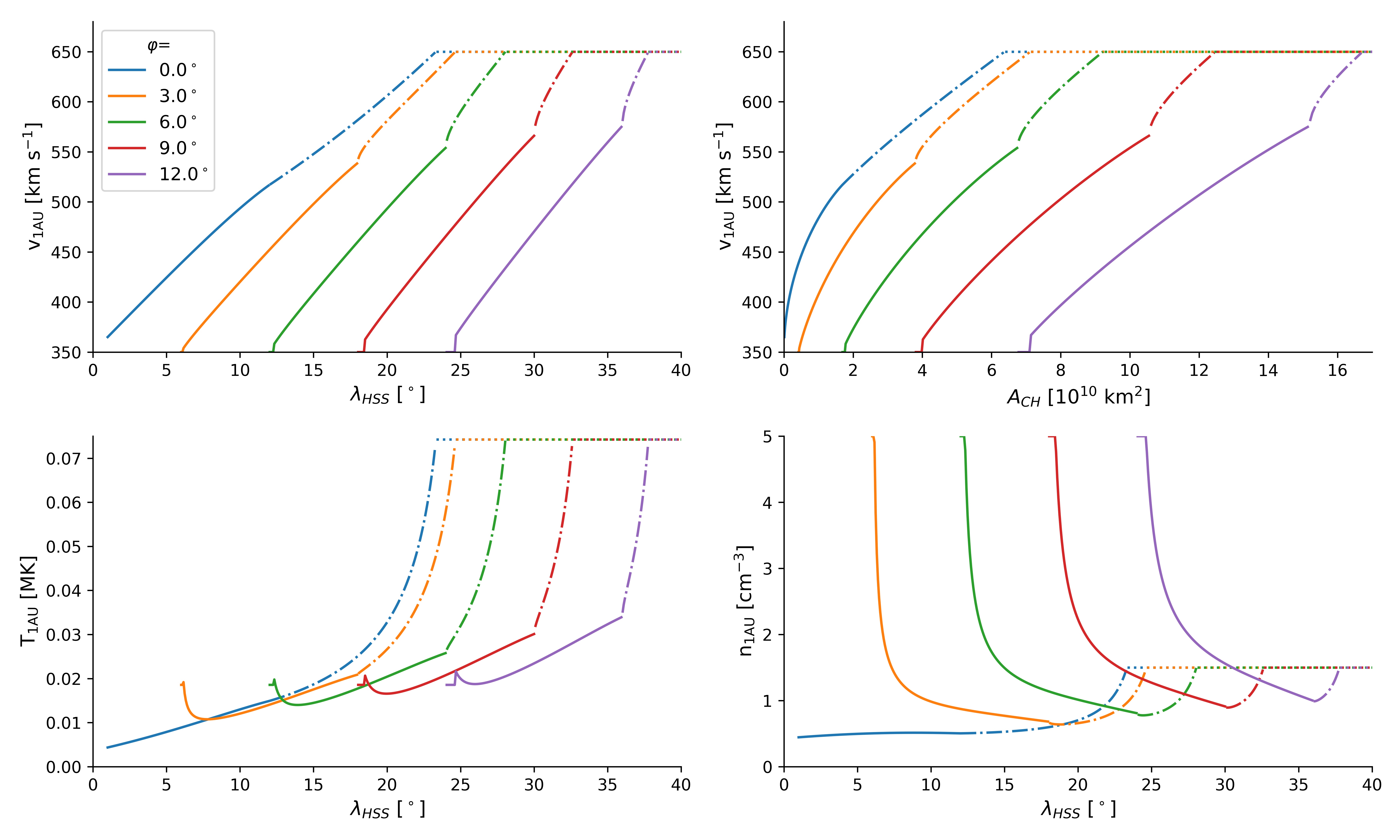}
\caption{Peak velocity of HSSs at \SI{1}{AU} versus their longitudinal widths (top left), and versus the areas of their source coronal holes at the solar surface (top right). Temperatures (bottom left) and densities (bottom right) of the HSS plasma parcels having the peak velocity at \SI{1}{AU}, versus the longitudinal widths of the HSSs. The data were derived for latitudinal displacements of \SI{0}{\degree}, \SI{3}{\degree}, \SI{6}{\degree}, \SI{9}{\degree}, and \SI{12}{\degree} to the center of the HSS.}
\label{fig:theory42}
\end{figure}

Here, we show how the properties of HSSs at \SI{1}{AU} depend on each other, and on the geometry of the HSS close to the Sun.
According to Equation \ref{eq:vat1au}, the peak velocity of HSSs is  dependent on the longitudinal width of the HSS close to the Sun. The longitudinal width of the HSS determines the spatial distance between the leading edge of the HSS, which coincides with the stream interface at this distance, and the trailing boundary region of the HSS. Consequently, it also affects the time and the distance from the Sun at which a plasma parcel originating from the trailing boundary region catches up with the stream interface. The larger the distance between the leading edge and trailing boundary region, the fewer plasma parcels from the trailing boundary region have impinged into the stream interface at a given distance from the Sun, and the higher is the peak velocity of the HSS at this distance. This explains the dependence of the peak velocity of HSSs on the longitudinal widths of the HSS. 
In the top left panel of Figure \ref{fig:theory42}, we show the peak velocities of HSSs at a distance of \SI{1}{AU} versus the longitudinal width of HSSs close to the Sun, at latitudinal displacements of \SI{0}{\degree}, \SI{3}{\degree}, \SI{6}{\degree}, \SI{9}{\degree}, and \SI{12}{\degree} to the HSS center. Dotted lines are associated with plasma parcels that originate from the HSS core region. Dashed-dotted lines represent plasma parcels originating from the trailing boundary region within a slice that crosses the HSS core region. And solid lines describe plasma parcels that originate from a slice that lies entirely in the northern or southern boundary region. The peak velocities are roughly linearly dependent on the longitudinal width of the HSSs and have similar slopes for all latitudinal displacements. Solely the x-intersections vary. The x-intersection gives the minimal latitudinal extent of a HSS that is needed so that the HSS crosses the longitudinal slice, and thus that a HSS appears in the data. 

In the top right panel of Figure \ref{fig:theory42}, we show the HSS peak velocities versus the areas of their solar source regions, i.e., coronal holes. The areas are thereby derived by $A_{CH} = \pi\ (\lambda_{HSS} /2)^2 $ in square degrees, and subsequently projected to a sphere at \SI{1}{R_{sun}} to derive their areas in \si{km^2}. This associates the areas of the HSSs with the areas of coronal holes at the solar surface. This functional relationship between longitudinal width and area results in that the lines get slightly skewed, dependent on the latitudinal displacement $\varphi$. 

In the bottom panels of Figure \ref{fig:theory42}, we show the dependence of the temperature and density of the plasma parcels at \SI{1}{AU} that are associated with the peak velocity versus the longitudinal width of the HSS. With increasing longitudinal width, the temperature increases and the density decreases, and finally saturates. The overshoot in the density decrease is related to the radial expansion of plasma parcels originating from the trailing HSS boundary.  Presuming a less steep sigmoidal velocity function across the HSS boundary or a wider HSS boundary region, respectively, reduces the density overshoot.

Note, that the kinks in the relationships between the longitudinal widths and areas to the velocities (at velocities of \SIrange{550}{600}{km s^{-1}}), in the temperatures (at \SIrange{30}{50}{kK}), and in the densities (at about \SI{1}{cm^{-3}}) are artifacts due to the approximation we used for modeling the HSS boundary regions. Different sigmoidal functions and definitions of the size of the boundary region will lead to slightly different results for the strength of the kinks. In purely numerical solutions of our initial equations, the kinks are smoothed out.

\subsection{Interrelation of HSS properties at \SI{1}{AU}}

\begin{figure}[t!]
\centering
 \includegraphics[width=.9\textwidth]{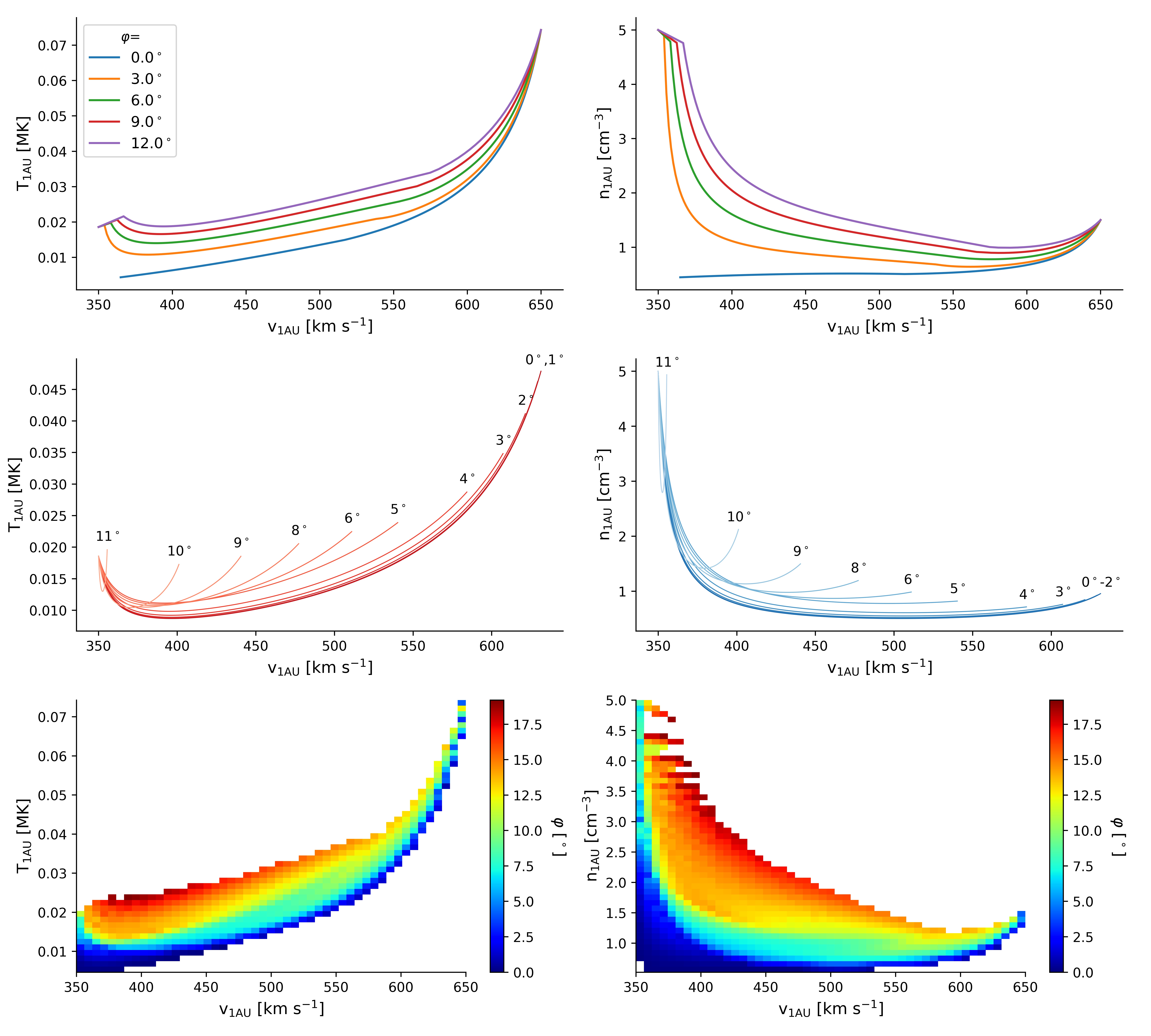}
\caption{Temperatures (left column) and densities (right column) versus the velocities of HSS plasma parcels for three subsets. First row: plasma parcels having the peak velocities of HSSs. Here, we considered all HSSs having a longitudinal widths of \SIrange{0}{40}{\degree}, at latitudinal displacements of the slice of \SI{0}{\degree}, \SI{3}{\degree}, \SI{6}{\degree}, \SI{9}{\degree}, and \SI{12}{\degree} to the center of the HSS. Second row: HSS plasma parcels originating from a HSS having a longitudinal widths of \SI{22}{\degree}, for latitudinal displacements of \SIrange{0}{11}{\degree} (labeled on top of the lines). Third row: all HSS plasma parcels, derived from all HSSs with longitudinal widths of \SIrange{0}{40}{\degree} at latitudinal displacements of \SIrange{0}{12}{\degree}.}
\label{fig:theory47}
\end{figure}

In this section, we show the interrelation between the velocity, density, and temperature of HSS plasma parcels at \SI{1}{AU}. We investigate the interrelation for three cases: 1) the distributions of plasma parcels that are related to the peak velocity of HSSs, 2) the distribution of HSS plasma parcels originating from a single longitudinal slice within a HSS, and 3) the statistical distribution of plasma parcels from large HSS datasets.

In the top panels of Figure \ref{fig:theory47}, we show the temperature and density of the plasma parcels representing the peak velocities at \SI{1}{AU} versus the peak velocities. For most of the distribution, the temperature of these plasma parcels increases and the densities decrease with increasing peak velocities. This is a consequence of their presumed (anti--) correlation close to the Sun, but amended by the radial expansion of the plasma parcels on their way to \SI{1}{AU}. Note, that for the longitudinal slice through the center of the HSS, i.e., at $\varphi = \SI{0}{\degree}$, we find small densities even for the small peak velocities. As the overall HSS has to be very small to provide small peak velocities in the center of HSS, and as thereby the chance that a satellite takes measurements right in the center of these HSSs becomes negligible, these low densities can probably be never observed. Therefore, we expect from observational HSS datasets, that analyze only the plasma parcels having the peak velocities, a correlation between the velocity and temperature and an anti-correlation between the velocity and density.

In the mid panels of Figure \ref{fig:theory47}, we show the temperature and density distributions of HSS plasma parcels versus their velocity for a HSS having a longitudinal width of \SI{22}{\degree}. Each line thereby represents plasma parcels originating from a different longitudinal slice through this HSS. For each slice, the dependence between the HSS density and temperature to the velocity has a unique curvature. This curvature is related to the radial expansion of the plasma parcels on their way from the Sun to \SI{1}{AU}, and thus to the velocity gradient across the trailing HSS boundary within that slice. The expansion is largest for plasma parcels originating from the center of the trailing boundary region, and less for parcels that arise from closer to western and eastern edge of the trailing boundary region, i.e., from the transition to the HSS core region or to the subsequent slow solar wind plasma. 

In the bottom panels of Figure \ref{fig:theory47}, we show statistical distributions of the temperatures and densities versus velocities, by superposing all HSS plasma parcels originating from all longitudinal slices for all HSSs up to a longitudinal width of \SI{40}{\degree}. The color gives the associated average latitudinal displacement of the origin of the plasma parcels to the center of the HSS. In this statistical dataset, the temperature increases with velocity, resulting from the original correlation of temperature and velocity across the HSS cross section close to the Sun. Again, the radial expansion of plasma parcels originating from the trailing boundary region results in additional cooling, which counteracts the temperature increase.
In contrast, the velocity-density distribution is much less confined. The decrease of density with increasing velocity originates from the anti-correlation of the density and velocity across the HSS cross section close to the Sun. Though, the radial expansion of plasma parcels additionally enhances the density decrease. Consequently, the densities can decrease very fast from slow solar wind to HSS values, dependent on the velocity gradient across the trailing boundary region, and thus strongly dependent on the latitudinal origin of the plasma parcels with respect to the  center of the HSS. This amplification for the density decrease increases the range at which densities appear at given velocities.

Note, that the curvatures in each of the plots depend on the radial expansion of plasma parcels from the Sun to Earth, and thus on the widths of the boundary region and the approximation we have used for the velocity profile across the boundary region. Larger widths, or less steep velocity gradients, respectively, result in smaller curvatures in these plots.

\subsection{Properties of stream interface at \SI{1}{AU}}
\begin{figure}[t!]
\centering
 \includegraphics[width=.9\textwidth]{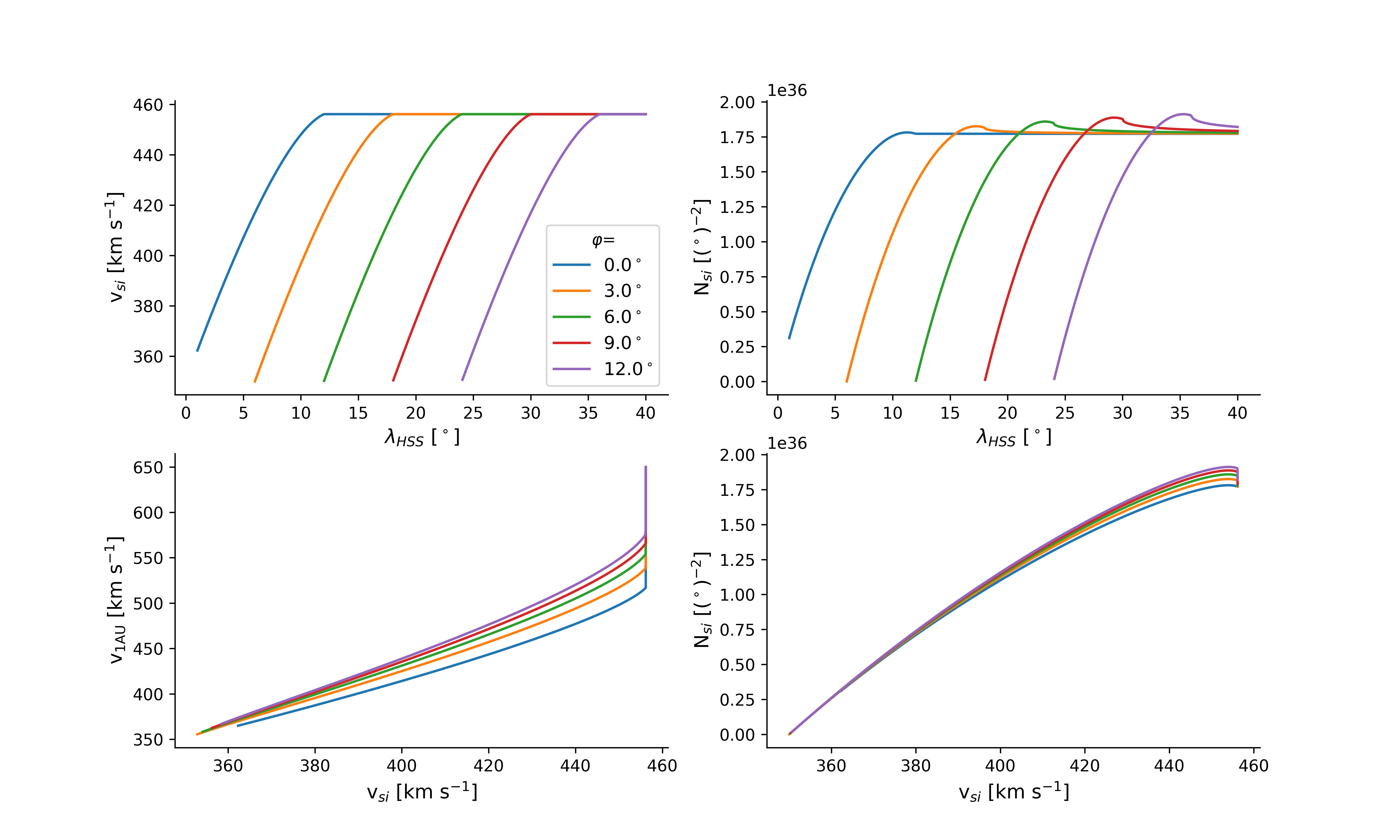}
\caption{Velocity of the stream interfaces at \SI{1}{AU} versus the longitudinal widths of HSSs (top left). Accumulated mass within a solid angle at the stream interface at \SI{1}{AU} versus the longitudinal widths of the HSSs (top right). Peak velocities of HSSs versus the velocities of the stream interfaces at \SI{1}{AU} (bottom left). Accumulated mass within a solid angle at the stream interfaces versus their velocities at \SI{1}{AU} (bottom right).}
\label{fig:theory44}
\end{figure}

Finally, we show how the properties of the stream interface at \SI{1}{AU} relate to the properties of HSSs at \SI{1}{AU}, and on the geometry of the HSSs close to the Sun.
The velocity of the stream interface depends on the momentum transferred from the HSS to the stream interface, and consequently on the underlying velocity profile along the longitudinal slice through the HSS. As the velocity profile depends on the longitudinal width of the HSS and the latitudinal displacement of the slice to the center of the HSS, the velocity of the stream interface is also dependent on these quantities, shown in the top left panel of Figure \ref{fig:theory44}. 
Furthermore, since both the peak velocity of the HSS and the velocity of the stream interface depend on the longitudinal width of the HSS, they are naturally correlated. However, it is important to note that this is a result of the mutual dependency on the longitudinal width of the HSS, and not a causal dependency. 

The accumulated mass of the stream interface is dependent on the velocity difference between the stream interface, the preceding slow solar wind, and the subsequent HSS (Eq. \ref{eq:theory_si}). A higher velocity of the stream interface implicates that less mass from the HSS accumulates, but also that more preceding slow solar wind plasma gets piled up. As the preceding slow solar wind plasma has a higher density than the HSS plasma, the slow solar wind term dominates. Consequently, the total accumulated mass increases with increasing velocity of the stream interface, which is shown in the bottom right panel of Figure \ref{fig:theory44}. 
Since the velocity of the stream interface is again dependent on the longitudinal width of the HSS, so is the accumulated mass of the stream interface as shown in the top panels of Figure \ref{fig:theory44}.

\subsection{Limitations} \label{sec:limitations}
In this study, a number of simplifications have been made so that the model was analytically solvable. These simplifications should not strongly affect the relationships between HSS properties derived throughout the study. However, some simplifications will affect the absolute values of the properties near Earth. 
 First, we have implicitly assumed that coronal holes extend radially into interplanetary space, forming the cross-section of HSSs at \SI{0.1}{AU}. In reality, each coronal hole has its unique expansion factor, and thus, the relationship between the area of coronal holes in the low corona to the area of the associated HSSs at \SI{0.1}{AU} is not trivial.
  Second, we have presumed that HSSs are propagating radially away from the Sun. HSSs are being bent towards the ecliptic very close to the Sun. Our calculations do not require radially propagating HSSs as assumed here for simplicity, but only that the HSS plasma propagates along a guiding line which does not affect its kinematics. Bent magnetic field lines provide such guiding lines for plasma, and thus do not affect our calculations regarding the HSS velocities. 
 Third, we have neglected a further acceleration of the solar wind in interplanetary space. Although this acceleration after \SI{0.1}{AU} is comparably small, it will still shift all velocities to slightly larger values. Fourth, we have assumed that the impingement of HSS plasma into the preceding slow solar wind plasma happens instantaneously, neglecting the extension of the SIR. This assumption might work well as long as the SIR has a limited extension compared to the spatial scales of the overall system; but it will break down at the latest when the backward wave steepens to a shock wave and decouples from the stream interface in the outer solar system. Fifth, we have neglected viscosity in the solar wind plasma, which results in a slower dispersion of the velocity distribution than presumed by our calculations. Sixth, our model is only one-dimensional for the propagation of plasma parcels from the Sun to Earth, and therefore neglects a possible redistribution of plasma along the two-dimensional SIR. This could affect the momentum balance between the HSS plasma and the preceding slow solar wind plasma at the stream interface, and thus the velocity of the stream interface and the amount of piled-up slow solar wind plasma.
 Seventh, the assumption of radially propagating HSSs, affect the temperatures we derive at Earth distance. MHD simulations show that HSSs originating from low and medium-latitude coronal holes usually are bent towards the ecliptic, whereby their expansion is inhibited towards higher latitudes. Thus, the plasma expands less into the latitudinal direction with distance to the Sun, which results in less cooling.
  Eighth, we have neglected waves in the plasma, which heat HSSs in interplanetary space, and could also stabilize HSS plasma, i.e., counteract velocity gradients. And ninth, adiabatic cooling by expansion, as presumed in our calculations, requires that work has to be done that transfers the energy away from the cooled gas. However, it is not clear to us what expansion work HSSs do during their propagation away from the Sun. This question is particularly interesting for plasma parcels in the HSS tail, as the preceding HSS plasma parcels are faster and thus run away, i.e., there should not be compression work involved. Note, that points seven and nine primarily affect the temperature; as the temperature is derived independently from the velocity and density in our calculations, this does not strongly affect the other main findings of our study.
 All above points will lead to deviations from the results presented here, and require more detailed studies. But they are not expected to change the physical explanation of how the area of solar coronal holes affects the solar wind properties near Earths.

\section{Summary, Discussion, and Conclusions} \label{sec:conclusions}

 In this study, we presumed that four main effects determine the properties of HSSs in interplanetary space at a given distance from the Sun. First, a velocity profile forms across the lateral HSS boundary close to the Sun. Due to the solar rotation, this spatial velocity profile maps into a temporal velocity profile into a given radial direction. Second, HSS plasma parcels propagate freely, i.e., with a constant radial velocity, in interplanetary space as long as they did not reach the SIR to the preceding slow solar wind. Third, the leading HSS plasma parcels impinges into the preceding slow solar wind plasma, thermalizes, and drives the SIR. And fourth, the characteristic properties of HSSs observed at a given distance from the Sun is given by the fastest HSS plasma parcels which did not impinge into the SIR yet.

By building an analytical model on these assumptions, we found that:
\begin{itemize}
\item The peak velocity of HSSs at Earth distance is dependent on the effective longitudinal widths of HSSs at \SI{0.1}{AU}, i.e., the length of a longitudinal slice through the HSS at \SI{0.1}{AU} that represents the foot points of the HSS plasma seen at Earth. The shorter the effective longitudinal widths, the more plasma from the HSS core region and trailing boundary region has already impinged into the stream interface at \SI{1}{AU}. The more `fastest' plasma parcels from the trailing boundary region have impinged into the stream interface, the smaller is the velocity of the leading plasma parcel, which did not yet impinge into the stream interface at \SI{1}{AU}, and thus the smaller is the peak velocity of the HSS seen at \SI{1}{AU}.
\item The lengths of the effective longitudinal width are dependent on the longitudinal widths of the overall HSS at \SI{0.1}{AU}, and the latitudinal displacement of the associated slice representing the foot points to the center of the HSS. When we approximate the longitudinal widths by its area (by assuming that the HSS has a circular cross-section), and further presume that the area of HSSs at \SI{0.1}{AU} statistically depends on the area of their source coronal holes, we can explain the long-known empirical relationship between the area of solar coronal holes and the peak velocities of the associated HSSs near Earth.
\item The temperature and density of HSS plasma parcels near Earth depend on their original temperature and density close to the Sun, and thus on their initial location within the HSS core or boundary region close to the Sun. 
\item Furthermore, their density and temperature near Earth depend on their expansion on their way from  Sun to Earth.
Plasma parcels originating from the boundary region of HSSs expand not only in the longitudinal and latitudinal direction, but also in the radial direction on their way from the Sun to Earth. The radial expansion thereby depends on the strengths of the radial velocity gradient within the parcels close to the Sun, which is again set by the initial location of the plasma parcels within the HSS core or boundary region, and in particular by the widths of the HSS boundary region.
\item As we presumed a correlation between the solar wind temperature and velocity close to the Sun, the radial expansion of the plasma parcels reducing the temperature counteracts this correlation. However, the velocity and temperature are still well correlated at Earth distance.
\item As we presumed an anti-correlation between the solar wind density and velocity close to the Sun, the radial expansion enhances the decrease of density with increasing velocity at larger distances from the Sun. This leads to a fast drop of density with increasing velocity in the associated scatter plot, and nearly destroys the initial anti-correlation between the solar wind density and velocity at Earth distance. 
\end{itemize}

In the following, we will shortly discuss some consequences of our results. 1) How our calculations help to understand from where within coronal holes the HSS plasma seen at Earth stems from, 2) how the correlation between HSS velocity, density, and temperature evolves from the Sun to Earth, and 3) how our calculations explain the empirical relationship between the HSS peak velocity at Earth and the area of coronal holes and why this relationship is similar to the distance to the coronal hole boundary model.

\subsection{From where within coronal holes does the solar wind plasma stem from that we see at Earth}

Based on our calculations, an extended temporal velocity plateau of HSSs at Earth indicates that not all plasma stemming from the HSS core close to the Sun has impinged into the stream interface yet. Consequently, the plasma from the plateau region at Earth stems from the inner region of coronal holes, away from the coronal hole boundaries, and the peak velocity of the HSS at Earth is similar to the peak velocity of the HSS close to the Sun. 
Suppose an extended plateau region at Earth is not apparent, i.e., that the temporal velocity profile is dominated by the tail during which the peak velocity decreases to slow solar wind velocities. This indicates that all the plasma from the HSS core region has already impinged into the stream interface. Consequently, the plasma seen at Earth stems mainly from the trailing boundary region of the coronal hole. The more plasma from the trailing boundary region has already impinged into the stream interface at Earth distance, the slower the peak velocity of the HSS at Earth will be. 
Thereby, the amount of plasma that has already impinged into the stream interface depends on the longitudinal extension of the HSS close to the Sun and the velocity gradient across the trailing boundary region. The velocity gradient again depends on the latitudinal distance of the foot points of the solar wind plasma in the coronal hole from its center. This interpretation goes along with \citet{nolte1977}, who studied eight HSSs at Earth distance in 1973. By ballistic back-mapping, they found that all the HSS plasma within the declining velocity tail at Earth distance stems from the trailing boundary region of the coronal holes.

Note, that these calculations have two interesting consequences. First, if almost all of the HSS plasma originating from a small- to medium-sized coronal hole has already impinged into the stream interface at a given distance from the Sun, we will see plasma that still has the composition of  HSSs, but only slow speeds. This ``slow Alfv\'enic solar wind" has been described by \citet{damicis2015} and \citet{stansby2020} as a distinct solar wind class, and they related their small velocities compared to usual HSSs to differing magnetic field geometries in the solar corona. We note, that some of these slow Alfv\'enic solar wind streams could also be usual HSSs where almost all of the plasma of the coronal hole has already impinged into the stream interface.
Second, in the very rare case of very small velocity gradients across the longitudinal HSS boundary region, it can happen that none of the HSS plasma has yet impinged into the stream interface at Earth distance, i.e., that the stream interface has not even formed. This case most likely happens if the plasma stems from very close to the latitudinal edge of coronal hole, where also the velocity is close to the adjacent slow solar wind velocity. In this case, one could observe almost undisturbed plasma stemming from leading boundary regions of coronal holes even at Earth distance.

\subsection{Evolution of the solar wind speed, density, and temperature with distance from the Sun}

In our calculations, we have presumed a velocity, density, and temperature gradient across the HSS boundary region close to the Sun, dependent on the distance to the HSS boundary. Consequently, the pairs of these three parameters have been assumed to be (anti-) correlated to each other close to the Sun. Thereupon, we have determined the expansion of the plasma parcels with distance from the Sun by their longitudinal and latitudinal expansion, which is related to the spherical geometry, times their radial expansion, which is related to their radial dispersion due to radial velocity gradients. Finally, the temperature has been derived under the assumption of adiabatic cooling and the negligence of wave heating; these limitations on the temperature are discussed in a subsequent section.

We presumed that the radial plasma velocity across the HSS core region is roughly constant, and consequently the associated radial velocity gradient is roughly zero. When we follow an isolated plasma parcel originating from the HSS core region while it propagates away from the Sun, its volume does not expand in the radial but only in the longitudinal and latitudinal direction. Therefore, its density follows a $r^2$-law, and its temperature follows a $r^{2 (\gamma-1)}$.
In contrast, plasma parcels originating from the boundary region of HSSs inherit a radial velocity gradient. Therefore, these plasma parcels naturally expand also in the radial direction while propagating away from the Sun. The rate of radial expansion depends on the velocity gradient, and thus on their initial location within the HSS boundary. Therefore, its density decrease as a function of distance will in general not follow a $r^2$-law, and consequently the temperature also not a $r^{2 (\gamma-1)}$ law. Furthermore, note, that the radial expansion of these plasma parcels, in combination with keeping their mean velocity constant, implies that the mass flux is in general not conserved. 
Thus, the functional relationship of the density and temperature of individual plasma parcels with distance from the Sun, and whether the mass flux is a conserved quantity, depends on the initial location of the plasma parcels within the HSS close to the Sun. 

Furthermore, when one compares the plasma properties of HSSs at various distance from the Sun, usually the properties around the peak velocity, or within the velocity plateau, respectively, are used. With increasing distance to the Sun, the fastest plasma parcels will have impinged into the stream interface, and thus the peak velocity will be represented by another plasma parcel originating from closer to the HSS trailing edge. Consequently, we should be careful when we compare the properties of HSSs at various distances from the Sun and interpret the results physically using thermodynamics, as we possibly compare the properties of distinctly different plasma parcels within the HSS. Interpreting the evolution of HSS properties with simple thermodynamics as done by \citet{perrone2019} might work well for HSSs that have a clear, extended velocity plateau, i.e., where all plasma parcels originate from the HSS core region. However, if such an extended velocity plateau is not apparent, the plasma parcels analyzed at the various distances will originate from different locations within the trailing boundary region close to the Sun, and the method will inevitably fail.

Our calculations further explain reasonably well why the velocities of HSS plasma parcels at Earth distance are still well correlated to their temperatures but why they are only weakly anti-correlated to their densities \citep[e.g., ][]{neugebauer1966, burlaga1970a, burlaga1970b, ogilvie1985, xu2015}. The additional decrease of temperature by the radial expansion of plasma parcels counteracts the original presumed correlation between the velocities and temperatures close to the Sun. This introduces a slope into the temperature-velocity distribution at Earth distance, but the temperature-velocity distribution remains well-defined. 
In contrast, the radial expansion further reduces the densities of the plasma parcels and thereby enhances the decrease of densities with increasing velocities we presumed close to the Sun. This enhancement results in that the densities can drop very fast with increasing velocities to the minimum value, in the case that the associated radial velocity  gradients close to the Sun have been large. This fast drop, superposed with the regular slower decreases of plasma parcels having smaller velocity gradients, results in the worse-defined density-velocity distribution shown in Figure \ref{fig:theory47}.

\subsection{The empirical relationships of the solar wind speed at Earth to the area of coronal holes, to the distance to the coronal hole boundary, and to the flux tube expansion factor}

The empirical relationship between the areas of solar coronal holes and the peak velocities of HSSs near Earth, as found by \citet{nolte1976, robbins2006, vrsnak2007, abramenko2009, karachik2011, rotter2012, hofmeister2018, garton2018, heinemann2020}, consists of three parts: 1) a roughly linear relationship between the areas of coronal holes and the peak velocities near Earths for small- to medium-sized coronal holes, 2) a saturation of the peak velocities of large coronal holes, and 3) smaller peak velocities for coronal holes of a given area located at higher latitudes.

In our calculations, we explained the roughly linear relationship between the areas of solar coronal holes and the peak velocities of the associated HSSs near Earth as an effect of the reduction of peak velocities by the continuous absorption of the fastest HSS plasma parcels by the stream interface, combined with the decreasing starting velocities of HSS plasma parcels across the HSS boundary region close to the Sun. After all the plasma originating from the leading boundary region and the HSS core region has impinged into the stream interface, the fastest plasma parcel originating from the trailing boundary region, which has not yet impinged into the stream interface, determines the peak velocity. The amount of plasma that has not yet impinged into the stream interface depends on the longitudinal widths of the coronal holes, or on their areas, respectively, and on the displacement of the longitudinal slice to the center of the coronal hole.
Further, the saturation of peak velocities near Earth for very elongated coronal holes arises from the fact that for very elongated coronal holes not all plasma parcels originating from the HSS core region have impinged into the stream interface at Earth distance. Thus, the peak velocities are given by plasma parcels originating from the HSS core region, which all have similar velocities. 
Finally, a larger latitudinal separation angle between the satellite measuring the HSS properties at Earth distance to the center of the coronal hole results in that we measure the HSS properties farther in its latitudinal flanks. Farther in the latitudinal flanks, the effective longitudinal widths of the HSS are reduced compared to the overall longitudinal widths of the HSS in its center. A smaller effective longitudinal width is again related to that more plasma from the trailing boundary region has already impinged into the stream interface, which results in smaller peak velocities measured at Earth distance.

Moreover, our results relate the empirical coronal hole area - HSS peak velocity relationship to the distance to the coronal hole boundary model \citep[DCHB;][]{riley2001} and the Wang-Sheeley-Arge model \citep[WSA;][]{arge2003}. The DCHB model, purely depending on the distance to the coronal hole boundary, and the WSA model, depending on the inverse flux tube expansion factor and the distance to the coronal hole boundary, are usually used to derive a HSS velocity, density, and temperature profile above coronal holes at \SI{0.1}{AU}. Both have in common that the velocities are large in the center of the HSS, and that they decrease to the ambient slow solar wind values at the HSS boundary. These are then used as input for MHD simulations to propagate the HSSs from the Sun to Earth. In this study, we have shown that the empirical relationship between the area of coronal holes and the peak velocities near Earth has implicitly incorporated the formation of the velocity profile above the coronal hole close to the Sun and the propagation phase from the Sun to Earth. The main difference in these three models is the presumed physical origin of the velocity profile close to the Sun. In the DCHB model, the acceleration of HSS plasma is presumed to depend on the distance to the coronal hole boundary, which is interpreted by diffusion and/or loop-opening processes at the coronal hole boundary. In the WSA model, the solar wind acceleration depends on the distance to the coronal hole boundary and a modeled flux tube expansion factor, whereby flux tubes near the boundary of coronal holes naturally show a larger expansion. A larger expansion is again interpreted with slower solar wind velocities. Within this study on the relationship between the coronal hole areas and HSS peak velocities at Earth, it does not matter where, or by which physical processes, the HSS boundary region forms; it is only important that it exists. Despite the different interpretations involved, all of these models describe a similar idea: that a velocity profile above the coronal hole forms, which sets the HSS properties close to the Sun, and that these HSSs are propagated from the Sun to Earth, taking the interaction with the preceding slow solar wind into account. 
\\[.5cm]
Finally, we like to add that the ongoing Parker Solar Probe and Solar Orbiter missions \citep{fox2016, mueller2020} will soon allow us to confront our model with more detailed observations. Parker Solar Probe's measurements during its perihelions close to the Sun will give us more insight into the distribution of the HSS properties across their cross-section near the Sun, while Solar Orbiter's unique orbit will allow us to analyze the radial evolution of HSSs and the associated SIRs in the inner heliosphere. These measurements will further improve our understanding on how HSSs evolve in the inner heliosphere, and how correspondingly the geometry of coronal holes on the Sun affects the properties of HSSs near Earth.

\begin{acknowledgements} 
We thank Lars Berger for discussions on the expansion of high-speed streams in interplanetary space and the associated cooling mechanisms.
Stefan J. Hofmeister is grateful for the Marietta-Blau fellowship awarded by the Austrian mobility service. This fellowship enabled him to visit the research groups at DTU Space, the University of Kiel, and KU Leuven for a total period of 10 months, and ultimately to achieve the results presented here. Furthermore, Stefan J. Hofmeister acknowledges support by the NASA Living with a Star program grant 80NSSC20K0183 and DFG grant 448336908. 
The project EUHFORIA 2.0 has received funding from the European Union’s Horizon 2020 research and innovation program under grant agreement No 870405. These results were also obtained in the framework of the projects
C14/19/089  (C1 project Internal Funds KU Leuven), G.0D07.19N  (FWO-Vlaanderen), SIDC Data Exploitation (ESA Prodex-12), and Belspo project B2/191/P1/SWiM.
Eleanna Asvestari acknowledges the support of the Academy of Finland, Project TRAMSEP, Academy of Finland Grant 322455.
Evangelia Samara was supported by a PhD grant awarded by the Royal Observatory of Belgium. Bojan Vr\v{s}nak acknowledges the support by the Croatian Science Foundation under the projects IP-01-2018-7549 (MSOC) and IP-2020-02-9893 (ICOHOSS).
\end{acknowledgements}

\end{document}